\pdfoutput=1
\newcommand*{\ATLASLATEXPATH}{}
\documentclass[cernpreprint, texlive=2016, UKenglish]{\ATLASLATEXPATH atlasdoc}
 
\usepackage[biblatex=true,backend=biber,subfigure]{\ATLASLATEXPATH atlaspackage}
\usepackage{\ATLASLATEXPATH atlasbiblatex}
 
\usepackage{\ATLASLATEXPATH atlasphysics}
\usepackage{\ATLASLATEXPATH atlasprocess}

\graphicspath{{logos/}{figures/}}
 
\usepackage{ANA-STDM-2018-04-PAPER-defs}

\AtlasTitle{Measurement of the \zllg production cross-section in \pp collisions at $\sqrt{s} =13~\TeV$ with the ATLAS detector}

\AtlasAbstract{
The production of a prompt photon in association with a \Zboson boson is studied in proton--proton collisions at a centre-of-mass energy $\sqrt{s} =$~13\,\TeV.
The analysis uses a data sample with an integrated luminosity of 139\,\ifb\ collected by the ATLAS detector at the LHC from 2015 to 2018.
The production cross-section for the process $\pp\to\lplmg+X$ ($\ell = e, \mu$) is measured within a fiducial phase-space region
defined by kinematic requirements on the photon and the leptons, and by isolation requirements on the photon.
An experimental precision of 2.9\% is achieved for the fiducial cross-section.
Differential cross-sections are measured as a function of each of six kinematic variables characterising the \lplmg system.
The data are compared with theoretical predictions based on next-to-leading-order and next-to-next-to-leading-order perturbative QCD calculations.
The impact of next-to-leading-order electroweak corrections is also considered.
}
 
\author{The ATLAS Collaboration}
 
\AtlasRefCode{STDM-2018-04}

\PreprintIdNumber{CERN-EP-2019-228}
 
\AtlasJournal{JHEP}
 
\AtlasJournalRef{JHEP 03 (2020) 054}
\AtlasDOI{10.1007/JHEP03(2020)054}
\hypersetup{pdftitle={ATLAS document},pdfauthor={The ATLAS Collaboration}}
 
\begin{document}
 
\maketitle
 
\tableofcontents
 
\section{Introduction}
\label{sec:intro}
Measurements of \Zboson boson production in association with a photon in high-energy collisions
provide tests of the electroweak sector of the Standard Model (SM) and can be used to search for new physics effects such as direct couplings of \Zboson bosons to photons.
Studies carried out at the Large Hadron Collider (LHC)
by the ATLAS~\cite{STDM-2012-07,STDM-2014-01} and CMS~\cite{CMS-EWK-11-009,CMS-SMP-12-020,CMS-SMP-13-014,CMS-SMP-14-019} collaborations
in proton--proton (\pp) interactions at centre-of-mass energies, $\sqrt{s}$, of 7\,\TeV\ and 8\,\TeV, as well as earlier measurements
from experiments at LEP~\cite{Achard:2004ds,Abdallah:2007ae,Abbiendi:2000cu} and the Tevatron~\cite{Abazov:2011qp,Abazov:2009cj,Aaltonen:2011zc}
in \ee and $\pbar p$ collisions,
have revealed no evidence for the existence of anomalous neutral gauge-boson interactions.
Measurements of \zg production rates in hadron--hadron collisions are also of interest, due to their sensitivity to higher-order effects predicted by perturbative QCD (pQCD).
A reliable characterisation of the properties of SM \zg production is of importance in searches for the decay $H\to\zg$ of the Higgs boson~\cite{HIGG-2016-14,CMS-HIG-17-007},
and in searches for other resonances in the \zg channel~\cite{HIGG-2016-14, CMS-EXO-17-005}, where non-resonant \zg production represents the dominant background process.
 
From 2015 to 2018 (\RunTwo), the LHC operated at a centre-of-mass energy of $\sqrt{s}=13~\TeV$.
The ATLAS Collaboration used the early part of the \RunTwo dataset, corresponding to an integrated luminosity of 36.1\,\ifb,
to measure the \zg production rate in the $\nu\overline{\nu}\gamma$~\cite{STDM-2017-18} and $b{\bar b}\gamma$~\cite{STDM-2017-14} channels,
in phase-space regions with photon transverse energy,\footnote{ATLAS uses a right-handed coordinate system
with its origin at the nominal interaction point (IP) in the centre of the detector and the $z$-axis along the beam pipe.
The $x$-axis points from the IP to the centre of the LHC ring, and the $y$-axis points upward.
Cylindrical coordinates $(r,\phi)$ are used in the transverse plane, $\phi$ being the azimuthal angle around the $z$-axis.
The transverse energy is defined as $\ET=E\sin\theta$, where $E$ is the energy and $\theta$ is the polar angle.
The pseudorapidity is defined as $\eta=-\ln\tan(\theta/2)$.
Angular separation is expressed in terms of $\Delta R \equiv \sqrt{(\Delta\eta)^{2} + (\Delta\phi)^{2}}$.}
\etg, greater than 150\,\GeV\ and 175\,\GeV, respectively.
The analysis of the neutrino channel allowed improved limits to be placed on anomalous $\Zboson\Zboson\gamma$ and $\Zboson\gamma\gamma$ couplings
which can arise in extensions of the SM~\cite{Gounaris:2000tb}.
The analysis presented here uses the full ATLAS \RunTwo dataset, with an integrated luminosity of 139\,\ifb,
to measure the \zg production cross-section
for events in which the \Zboson boson decays into an electron or muon pair, \Zlplm $(\ell = e,\mu)$.
Compared with the neutrino channel, the \lplmg channel allows cross-section measurements to be made over a wider range of \etg and with lower background, but with reduced sensitivity to anomalous gauge-boson couplings~\cite{STDM-2014-01,STDM-2015-21}.
 
Inclusive samples of \epemg and \mpmmg events are selected and used to measure the \zg production cross-section
within a fiducial phase-space region defined by the kinematic properties of the lepton pair and the photon,
including a requirement that the invariant mass, \mll, of the \leplep pair be greater than 40\,\GeV\,
and that the sum, $\mll + \mllg$, of the invariant masses of the lepton pair and the \lplmg system be greater than 182\,\GeV.
The latter requirement ensures that the measurement is dominated by events in which the photon is emitted from an initial-state quark line in the hard-scattering process,
as in \Fig{\ref{fig:feynmanZg}(a)}, rather than from a final-state lepton, as in \Fig{\ref{fig:feynmanZg}(b)}.
The \mll distribution for selected \lplmg events thus displays a dominant resonant peak centred on the \Zboson boson mass,
above a smaller, non-resonant component due to the presence of virtual photon exchange.
The contribution from events in which the selected photon is produced from the fragmentation of a quark or a gluon, as illustrated in \Figs{\ref{fig:feynmanZg}(c)}{\ref{fig:feynmanZg}(d)},
is suppressed experimentally by requiring that the photon be unaccompanied by significant activity from other particles in the event (isolation),
and removed theoretically by imposing smooth-cone isolation criteria on the photon at parton level~\cite{Frixione:1998jh}.
 
\begin{figure}[hbtp]
\begin{center}
\includegraphics[width=0.30\textwidth]{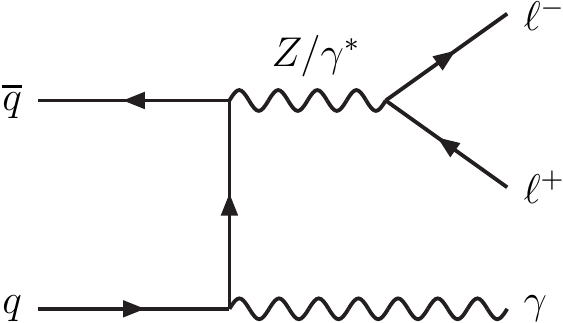}  \qquad
\includegraphics[width=0.30\textwidth]{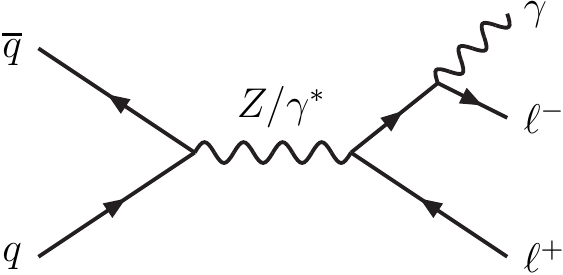}  \\
(a) \hspace*{50mm} (b)  \\
\includegraphics[width=0.30\textwidth]{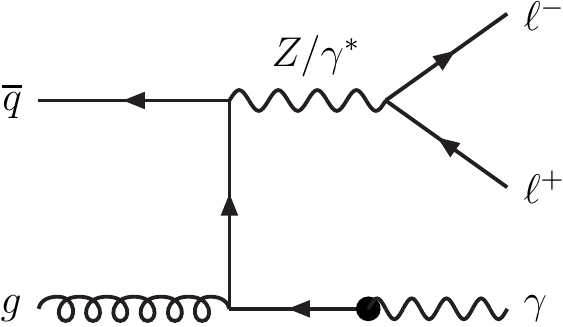}  \qquad
\includegraphics[width=0.30\textwidth]{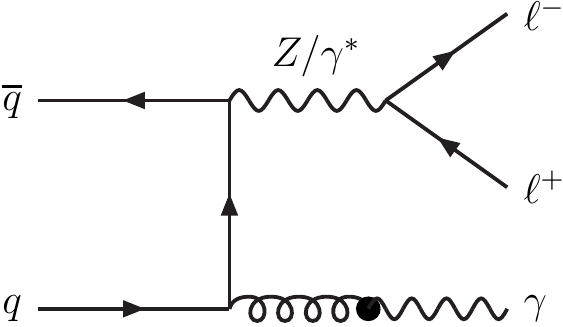}  \\
(c) \hspace*{50mm} (d)  \\
\caption{Feynman diagrams for \lplmg production:
(a) initial-state photon radiation from a quark line; (b) final-state photon radiation from a lepton;
and (c,d) contributions from the $\Zboson+q(g)$ processes in which a photon is produced from the fragmentation of a quark or a gluon.
}
\label{fig:feynmanZg}
\end{center}
\end{figure}
 
The measurements of the rate and kinematic properties of \zg production in the fiducial phase-space region
are compared with SM predictions obtained from parton-level calculations carried out in pQCD at next-to-leading order (NLO) and
next-to-next-to-leading order (NNLO) in the strong coupling constant \alphas,
as well as with predictions from parton shower Monte Carlo (MC) event generators with leading-order (LO) and NLO matrix elements.
The effect of NLO electroweak (EW) corrections on the predictions at NNLO in pQCD is also considered.
A small contribution to \zg production arises from the vector-boson scattering process $pp\to\zgjj$~\cite{STDM-2017-26, CMS-SMP-14-018}, and is considered to be part of the signal.
Differential cross-sections are measured as functions of the transverse energy, \etg, and absolute pseudorapidity, \aetag, of the photon,
and as functions of the invariant mass, \mllg, and transverse momentum, \ptllg, of the \lplmg system, the ratio \rllg,
and the angle, \dphillg, between the transverse directions of the $\ell^+\ell^-$ pair and the photon.
Differential cross-sections in the latter three variables have not been measured previously for \zg
production, and provide particularly sensitive tests of higher-order pQCD calculations.

\section{The ATLAS detector}
\label{sec:detector}
The ATLAS experiment~\cite{PERF-2007-01} at the LHC is a multipurpose particle detector with a forward--backward symmetric cylindrical geometry and nearly $4\pi$ coverage in solid angle.
Its major components are an inner tracking detector (ID) surrounded by a thin superconducting solenoid providing a 2\,T axial magnetic field,
electromagnetic (ECAL) and hadron (HCAL) calorimeters, and a muon spectrometer (MS).
The ID is composed of a silicon pixel detector (including the insertable B-layer~\cite{PIX-2018-001,ATLAS-TDR-2010-19} installed before the start of \RunTwo)
and a silicon microstrip tracker (SCT), both of which cover the pseudorapidity range $|\eta|<2.5$,
together with a transition radiation tracker (TRT) with an acceptance of $|\eta|<2.0$.
The TRT provides identification information for electrons by the detection of transition radiation.
The MS is composed of three large superconducting air-core toroid magnets, a system of three stations of chambers for tracking measurements,
with high precision in the range $|\eta|<2.7$, and a muon trigger system covering the range $|\eta|<$ 2.4.
 
The ECAL is composed of alternating layers of passive lead absorber interspersed with active liquid-argon (LAr) gaps and covers the pseudorapidity range $|\eta| < 3.2$.
For $|\eta| < 2.5$ the calorimeter is segmented longitudinally in shower depth into three layers, with the first layer having the highest
granularity in the $\eta$ coordinate, and the second layer collecting most of the electromagnetic shower energy.
A thin presampler layer precedes the ECAL over the range $|\eta|<1.8$, and is used to correct for energy loss upstream of the calorimeter.
The HCAL, surrounding the ECAL, employs either scintillator tiles or LAr as the active medium, and either steel or copper as the absorber material.
Two copper/LAr and tungsten/LAr forward calorimeters extend the acceptance up to $|\eta| = 4.9$.
 
Collision events are selected using a two-level trigger system~\cite{TRIG-2016-01}.
The first-level trigger is implemented in custom electronics and, using a subset of the information from the detector,
reduces the trigger rate to about 100\,kHz from the original 40\,MHz LHC proton bunch-crossing rate.
The second-level trigger is a software-based system which runs algorithms similar to those implemented in the offline reconstruction software, yielding a recorded event rate of about 1\,kHz.
\section{Data and simulated event samples}
\label{sec:samples}
The data used in this analysis were collected in proton--proton collisions at $\sqrt{s} = 13~\TeV$ from 2015 to 2018.
After applying criteria to ensure good ATLAS detector operation, the total integrated luminosity useful for data analysis is 139\,\ifb.
The uncertainty in the combined 2015--2018 integrated luminosity is 1.7\%~\cite{ATLAS-CONF-2019-021}, obtained using the LUCID-2 detector~\cite{LUCID2} for the primary luminosity measurements.
The average number of inelastic \pp interactions produced per bunch crossing for the dataset considered is $\meanmu = 33.7$.
 
Simulated event samples are used to correct the signal yield for detector effects and to estimate several background contributions.
The simulated samples were produced with various MC event generators, processed through a full ATLAS detector simulation~\cite{SOFT-2010-01}
based on \textsc{Geant4}~\cite{Agostinelli:2002hh}, and reconstructed with the same software as used for the data.
All MC samples are corrected with data-driven correction factors
to account for differences in photon and lepton trigger, reconstruction, identification and isolation performance between data and simulation.
Additional \pp interactions (pile-up) occurring in the same and neighbouring bunch crossings were modelled
by overlaying each MC event with minimum-bias events generated using \PYTHIAV{8.186}~\cite{Sjostrand:2007gs} with the A3 set of tuned parameters~\cite{ATL-PHYS-PUB-2016-017}
and the \NNPDFVV{2.3}{LO}~\cite{Ball:2012cx} set of parton distribution functions (PDFs).
The MC events were then reweighted to reproduce the distribution of the number of \pp interactions per bunch crossing observed in the data.
 
\begin{table}[t]
\caption{Summary of simulated MC event samples for the \lplmg signal process (first two rows) and for various background processes (lower six rows).
The third and fourth columns give the pQCD order and the PDF set used in the hard-scattering matrix element calculations.
The rightmost column specifies the generator used to model parton showering, hadronisation, the underlying event and multiple parton interactions.
}
\label{table:MCsamples}
 
\begin{center}
\begin{tabular}{lllll}
\hline
Process                 &     Generator           &  Order  &  PDF Set          &  PS/UE/MPI               \\
\hline
\llg                    &  \SHERPAV{2.2.4}        &  LO     &  \NNPDFVV{3.0}{NNLO}  &  \SHERPAV{2.2.4}         \\    
\llg                    &  \MGMCatNLOV{2.3.3}     &  NLO    &  \NNPDFVV{3.0}{NLO}   &  \PYTHIAV{8.212}         \\    
\hline
\Zjets                  &  \POWHEGBOXV{v1}             &  NLO    &  CT10 NLO             &  \PYTHIAV{8.186}         \\    
\ttg                    &  \MGMCatNLOV{2.3.3}     &  LO     &  \NNPDFVV{2.3}{LO}    &  \PYTHIAV{8.212}         \\    
\wz, \zz                &  \SHERPAV{2.2.2}        &  NLO    &  \NNPDFVV{3.0}{NNLO}  &  \SHERPAV{2.2.2}         \\    
\wwg, \wzg              &  \SHERPAV{2.2.5}        &  NLO    &  \NNPDFVV{3.0}{NNLO}  &  \SHERPAV{2.2.5}         \\    
$\tau\tau\gamma$        &  \SHERPAV{2.2.4}        &  LO     &  \NNPDFVV{3.0}{NNLO}  &  \SHERPAV{2.2.4}         \\    
$H\to\zg$               &  \POWHEGBOXV{v2}             &  NLO    &  PDF4LHC15 NNLO       &  \PYTHIAV{8.212}         \\    
\hline
\end{tabular}
\end{center}
\end{table}
 
Samples of simulated \epemg and \mpmmg events with lepton-pair invariant mass greater than 10\,\GeV\ generated using \SHERPAV{2.2.4}~\cite{Bothmann:2019yzt}
with the \NNPDFVV{3.0}{NNLO}~\cite{Ball:2014uwa} PDF set
are used to estimate the effects of detector efficiency and resolution on the expected number of signal events.
These samples were generated including all Feynman diagrams with three electroweak couplings, with up to three additional final-state partons at LO in pQCD,
and merged with the \SHERPA parton shower~\cite{Schumann:2007mg} according to the \MEPSLO prescription~\cite{Hoeche:2011fd, Hoeche:2012yf, Catani:2001cc, Hoeche:2009rj}.
For studies of systematic uncertainties, an alternative signal sample was produced
using the generator \MGMCatNLOV{2.3.3}~\cite{Alwall:2014hca} with up to three additional final-state partons, where up to one additional final-state parton is at NLO accuracy,
and using the \NNPDFVV{3.0}{NLO} PDF set.

The dominant background to the \zg signal, arising from events containing a \Zboson boson together with associated jets in which one of the jets is misidentified as a photon,
is estimated using a data-driven method.
To validate the method and to estimate the associated systematic uncertainties,
a simulated sample of \Zjets events (with $\Zboson\to ee$ or $\Zboson\to\mu\mu$) was produced.
The sample was generated with \POWHEGBOXV{v1}~\cite{Nason:2004rx, Frixione:2007vw, Alioli:2010xd, Alioli:2008gx} at NLO accuracy,
using the CT10~\cite{Lai:2010vv} NLO PDF set.

Background contributions from $\ell\nu\ell\ell$ (`\wz'), $\ell\ell\ell\ell$ (`\zz'), \wwg and \wzg production
(including decays of the \Wboson or \Zboson boson to final states involving a $\tau$-lepton)
are estimated from simulated event samples generated using the \SHERPAV{2.2.2} (\wz, \zz) or \SHERPAV{2.2.5} (\wwg, \wzg) generators,
using the \MEPSNLO prescription~\cite{Hoeche:2011fd, Hoeche:2012yf, Catani:2001cc, Hoeche:2009rj},
and using the \OPENLOOPS library~\cite{Cascioli:2011va, Denner:2016kdg} to provide the virtual QCD corrections to matrix elements at NLO accuracy.
The background contribution from $\tau^+\tau^-\gamma$ production is estimated from a simulated event sample
generated using \SHERPAV{2.2.4} with the same LO configuration as used to generate the \SHERPA signal sample described above.
The background from top-quark production is estimated from a simulated sample of \ttg events as used in \Ref{\cite{TOPQ-2017-14}}, with one or both of the top quarks decaying semileptonically,
generated with \MGMCatNLOV{2.3.3} at LO with the \NNPDFVV{2.3}{LO} PDF set.
The background from events containing $H\to\Zboson\gamma$ decays (with $\Zboson\to ee$ or $\Zboson\to\mu\mu$)
is estimated using a simulated event sample as used in \Ref{\cite{HIGG-2016-14}} generated with \POWHEGBOXV{v2},
using the MiNLO~\cite{Hamilton:2015nsa} and NNLOPS~\cite{Hamilton:2013fea} approaches,
and using the PDF4LHC15 NNLO PDF set~\cite{Butterworth:2015oua}.
 
The \POWHEGBOX and \MGMCatNLO generators were interfaced to \PYTHIAV{8.186} and to \PYTHIAV{8.212}~\cite{Sjostrand:2014zea}, respectively,
for parton showering and hadronisation, and to model the underlying event and multiple parton interactions.
The \PYTHIA generator was configured using the A14 set of tuned parameters~\cite{ATL-PHYS-PUB-2014-021},
except for the simulated \Zjets and $H\to\Zboson\gamma$ samples generated with \POWHEGBOX where the AZNLO set of tuned parameters~\cite{STDM-2012-23} was used.
The \EVTGENV{1.2.0} and \EVTGENV{1.6.0} programs~\cite{Lange:2001uf} were used to describe the properties of bottom and charm hadron decays
in the samples generated using \POWHEGBOX and \MGMCatNLO, respectively,
and the \PHOTOS~\cite{Davidson:2010ew} generator was used for the simulation of photon bremsstrahlung in the decays of particles and resonances.
 
A summary of the signal and background MC samples used in the analysis is presented in \Tab{\ref{table:MCsamples}}.

For the generation of the \zg signal samples, and the $\tau\tau\gamma$, \wwg and \wzg background samples,
photon isolation criteria were imposed at parton level using the smooth-cone isolation prescription of \Ref{\cite{Frixione:1998jh}}.
This removes contributions in which the photon is produced from quark or gluon fragmentation (\Figs{\ref{fig:feynmanZg}(c)}{\ref{fig:feynmanZg}(d)})
in a way which is infrared safe to all orders of perturbation theory.
The smooth-cone isolation prescription considers a cone of variable opening angle \frixdel, with maximum opening angle \frixdelmax, centred around the photon direction,
and requires that the summed transverse energy of partons inside the cone is always less than a specified fraction of $\etg$.
This fraction has a maximum value \frixepsmax for a cone of maximum size $\frixdel = \frixdelmax$,
and tends smoothly to zero as $\frixdel\to 0$ according to the function $[(1-\cos\frixdel)/(1-\cos\frixdelmax)]^n$.
In all cases, the smooth-cone isolation parameters were set to the values $\frixdelmax = 0.1$, $\frixepsmax = 0.1$ and $n = 2$.
 
\section{Selection of \lplmg events}
\label{sec:selection}
Candidate \lplmg events are selected by requiring the presence of a photon with high $\etg$
together with an opposite-charge, same-flavour lepton (electron or muon) pair.
No explicit requirements are made on the presence or absence of other activity in the event, such as additional photons or leptons, or jets.
Background events from processes producing non-prompt photons or leptons are removed by imposing isolation requirements on the photon and the two leptons.
 
Event candidates in both data and MC simulation are required to have fired at least one unprescaled single-electron or single-muon trigger.
For data recorded in 2015, the lowest \pt threshold for such triggers was 24\,\GeV\ for electrons~\cite{TRIG-2018-05} and 20\,\GeV\ for muons~\cite{TRIG-2016-01}.
For data recorded during 2016--2018, due to the higher instantaneous luminosity, the lowest \pt trigger threshold for both the electrons and muons was raised to 26\,\GeV,
and tighter lepton isolation and identification requirements were imposed.
Triggers with higher \pT thresholds but with looser isolation or identification criteria were also used to increase the total data-taking efficiency.
The trigger efficiency for \lplmg events satisfying all the selection criteria described below is about 99\%.
This is determined using a simulated signal sample, corrected to reflect the trigger efficiencies measured in data
using correction factors determined in studies of $\Zboson\to\ell\ell$ decays.

\subsection{Photon and lepton selection}  \label{sec:objsel}
 
Photon and electron candidates are reconstructed~\cite{EGAM-2018-01} from clusters of energy deposits in the ECAL,
together with information about charged tracks reconstructed in the ID.
Photon clusters are required to have a pseudorapidity in the range $|\eta|<2.37$, and to have a transverse energy $\etg>30$\,\GeV.
Electron clusters with $\pT>25$\,\GeV\ are required to lie in the range $|\eta|<2.47$, and to be matched to a reconstructed track in the ID.
For both the photons and electrons, the transition region between the barrel and endcap regions (\mbox{$1.37<|\eta|<1.52$}) is excluded.
Photon candidates are classified either as \textit{converted}
(the photon cluster is matched to a reconstructed conversion vertex formed either from two oppositely charged tracks or from a single track consistent with having originated from a photon conversion)
or as \textit{unconverted} (matched to neither a conversion vertex nor an electron track).
Converted and unconverted photon candidates are both used in the analysis.
Muon candidates are reconstructed~\cite{PERF-2015-10} from tracks in the MS that are matched to a corresponding track in the ID.
The muon momentum is calculated by combining the MS measurement, corrected for the energy deposited in the calorimeters, and the ID measurement.
The \pt\ of the muon must be greater than 25\,\GeV\ and its pseudorapidity must satisfy $|\eta|<2.5$.
 
The shower shapes produced in the ECAL are used to identify photons and electrons.
Photons are required to satisfy all the requirements on shower shape variables which correspond to the \textit{Tight} photon identification criteria of \Ref{\cite{EGAM-2018-01}}.
The Tight photon identification efficiency ranges from 82--85\% for photons with $\etg \approx 30$\,\GeV\ to 90--98\% for $\etg > 100$\,\GeV,
depending on the pseudorapidity region of the detector and on the conversion status of the photon candidate.
Electrons are identified using a discriminant that is the value of a likelihood function constructed from quantities describing the shape of the electromagnetic shower in the calorimeter,
together with quantities characterising the electron track and the quality of the track--cluster matching~\cite{PERF-2017-01}.
Electron candidates are required to satisfy the \textit{Medium} likelihood requirement of \Ref{\cite{EGAM-2018-01}},
which provides an identification efficiency of about 80\% (93\%) for electrons of $\pt\approx 25$\,\GeV\ (100\,\GeV).
Muon candidates are required to satisfy the \textit{Medium} identification criteria of \Ref{\cite{PERF-2015-10}};
these include requirements on the numbers of hits matched to the tracks reconstructed in the ID and in the MS,
and on the probability of compatibility between the ID and MS momentum measurements.
The overall efficiency of the muon reconstruction and identification is about 97\%, with no strong dependence on the muon $\pt$.
 
Electron and muon candidates are required to originate from the primary vertex\footnote{Each primary vertex candidate is reconstructed
from at least two associated tracks with $\pT>0.4$\,\GeV.
The primary vertex is selected among the primary vertex candidates as the one with the highest sum of the squared transverse momenta of its associated tracks.}
by demanding that the significance of the transverse impact parameter,
defined as the absolute value of the track transverse impact parameter, $d_0$, measured relative to the beam trajectory, divided by its uncertainty, $\sigma_{d_{0}}$,
satisfy $|d_{0}|/\sigma_{d_{0}} < 3$ for muons and $|d_{0}|/\sigma_{d_{0}} < 5$ for electrons.
The difference $\Delta z_0$ between the value of the $z$ coordinate of the point on the track at which $d_0$ is defined,
and the longitudinal position of the primary vertex, is required to satisfy $|\Delta z_0\cdot \sin\theta| < 0.5$\,mm both for muons and electrons.
 
Photon, electron and muon candidates are required to be isolated from other particles.
In all cases, the isolation criteria place requirements on the sum, \ptiso, of the scalar transverse momenta of tracks with $\pt > 1$\,\GeV,
and on the sum, \etiso, of the transverse energy of topological clusters~\cite{PERF-2014-07},
within cones defined in terms of the distance $\Delta R$ to the photon or lepton.
The quantity \ptiso is computed using tracks which are matched to the primary vertex,
or which are not matched to any vertex but have a distance of closest approach to the primary vertex along the beam axis $|\Delta z_0\cdot\sin\theta| < 3$\,mm.
\stilltodo{Check this for muons.}
Tracks associated with the electron, muon or photon candidate are excluded from the track isolation \ptiso.
The calorimeter isolation \etiso is corrected on an event-by-event basis for the energy deposited by the photon or lepton candidate,
and, using the method described in Refs.~\cite{Cacciari:2008gn,Cacciari:2009dp, STDM-2010-08}, for the contribution from the underlying event and pile-up.
 
Photon candidates are required to satisfy the \textit{FixedCutLoose} isolation criteria of \Ref{\cite{EGAM-2018-01}}.
The FixedCutLoose isolation employs a cone of size $\Delta R = 0.2$ for both the track and calorimeter isolation,
and requires $\ptiso/\etg < 0.05$ and $\etiso/\etg < 0.065$.
Electron candidates are required to satisfy the \textit{FCLoose} isolation criteria of \Ref{\cite{EGAM-2018-01}}.
The track isolation \ptiso for electrons employs a cone of $\pt$-dependent size up to $\Delta R = 0.2$,
while the calorimeter isolation \etiso is computed using a cone of fixed size $\Delta R = 0.2$.
The FCLoose isolation for electrons requires $\ptiso/\pt < 0.15$ and $\etiso/\pt < 0.2$.
Muon candidates are required to satisfy the \textit{FCLoose\_FixedRad} isolation criteria of \Ref{\cite{PERF-2015-10}}.
The track isolation \ptiso for muons employs a cone of \pt-dependent size up to $\Delta R = 0.3$ ($\Delta R = 0.2$) for muons with transverse momentum less than (greater than) 50\,\GeV,
while the calorimeter isolation \etiso uses a cone of fixed size $\Delta R = 0.2$.
The FCLoose\_FixedRad isolation for muons requires $\ptiso/\pt < 0.15$ and $\etiso/\pt < 0.3$.
 
For unconverted (converted) photons, the isolation requirements have an efficiency of about 88\% (80\%) for photons with $\etg\approx 30$\,\GeV,
rising to about 98\% (96\%) for $\etg > 200$\,\GeV.
For leptons, the isolation requirements have an efficiency of about 98\%
(close to 100\%) for electrons or muons with $\pt\approx 25$\,\GeV\ ($\pt > 50$\,\GeV).
 
In addition to the isolation requirements above,
photon candidates are required to be separated from all electron and muon candidates in the event by $\Delta R(\ell,\gamma) > 0.4$,
and electron candidates are required to be separated from all muon candidates in the event by $\Delta R(\mu,e) > 0.2$.

\subsection{Signal region definition}  \label{sec:SRdef}
 
Candidate \lplmg signal events are selected by requiring that they contain at least one opposite-charge, same-flavour pair of lepton candidates and at least one photon candidate.
One of the electrons or muons in the lepton pair must be matched to the single-lepton trigger electron or muon which triggered the event.
One of the electrons or muons in the lepton pair must have $\pT > 30$\,\GeV.
The opposite-charge, same-flavour lepton pair with the highest summed lepton \pT (the {\textit{leading lepton pair}}) is selected.
The invariant mass \mll of the leading lepton pair is required to be greater than $40\,\GeV$, to remove contributions from low-mass resonances.
The \lplmg system is formed from the leading lepton pair and the highest-\etg photon candidate in the event.
To suppress events where the \lplmg system originates from the decay of a \Zboson,
events are selected by requiring the sum of \mll and the invariant mass \mllg of the \lplmg system to be greater than 182\,\GeV,
approximately twice the mass of the \Zboson boson~\cite{STDM-2015-21}.
The impact of this requirement on the selection of events in data is shown in \Fig{\ref{fig:diagonalcut}}.

\begin{figure}[ht]
\begin{center}
\includegraphics[width=0.7\textwidth]{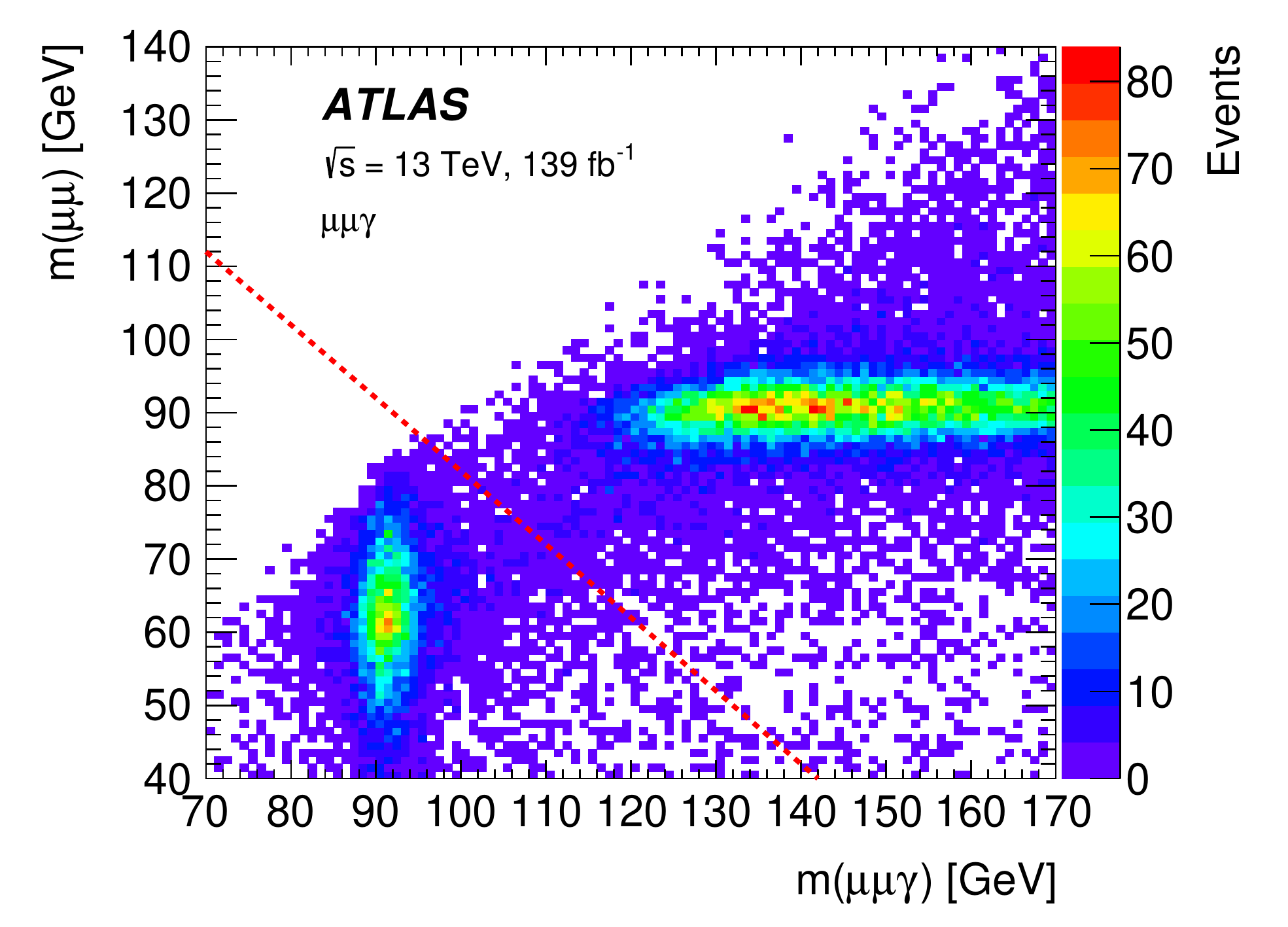}
\caption{Two-dimensional distribution of \mll and \mllg for events satisfying all \mpmmg selection criteria except that on the sum of \mll and \mllg.
The diagonal dashed line shows the selection $\mll+\mllg>182$\,\GeV\ used to ensure that the measurement
is dominated by events in which the photon is emitted from an initial-state quark.}
\label{fig:diagonalcut}
\end{center}
\end{figure}

The photon, lepton and event selection requirements above define the signal region (SR) and are summarised in \Tab{\ref{table:ZgSigReg}}.
After imposing all SR selection requirements, a total of 41343 \epemg events and 54413 \mpmmg events are selected in the data.

\begin{table}[t]
\caption{Definition of the \lplmg signal region.
The selection criteria for photons and leptons are presented in the upper part of the table, while the event-level selection criteria are presented in the bottom row.
For the lepton \pT requirements, the first (second) number specifies the minimum allowed \pT of the lepton with the highest (second-highest) value of transverse momentum.
}
\label{table:ZgSigReg}
\begin{center}
\begin{tabular}{lccc}
\hline
~~   &     Photons  & Electrons & Muons  \\
\hline
Kinematics:        &  $\ET > 30$~\GeV & $\pT>30, 25$~\GeV & $\pT > 30, 25$~\GeV  \\
~~   &  $|\eta| < 2.37$ & $|\eta| < 2.47$ & $|\eta| < 2.5$  \\
~~   &  ~excl.\ $1.37 < |\eta| < 1.52$~ & ~excl.\ $1.37 < |\eta| < 1.52$~ & ~~  \\
Identification:~~  &   Tight~\cite{EGAM-2018-01}  &  Medium~\cite{EGAM-2018-01}   &  Medium~\cite{PERF-2015-10}  \\
Isolation:         &   FixedCutLoose~\cite{EGAM-2018-01}  &  FCLoose~\cite{EGAM-2018-01}   &  FCLoose\_FixedRad~\cite{PERF-2015-10}  \\
~~   &    $\Delta R(\ell,\gamma) > 0.4$  &  $\Delta R(\mu,e) > 0.2$  &  ~~  \\
\hline
Event selection:   &   \multicolumn{3}{c}{$\mll > 40\,\GeV$, \hspace*{3mm} $\mll + \mllg > 182\,\GeV$}  \\
\hline
\end{tabular}
\end{center}
\end{table}

\section{Background estimation}
\label{sec:backgrounds}
The dominant source of background to the \zllg signal originates from \Zjets production in which a jet is misidentified as a photon.
Other, smaller, background contributions arise from top quark or multiboson production,
and from {\textit{pile-up background}} in which the selected photon and the selected lepton pair arise from different \pp interactions occurring within the same LHC bunch crossing.
The production of \zg pairs giving the final state $\tau\tau\gamma$ is considered to be a background process rather than part of the signal.
The \Zjets and pile-up backgrounds are estimated using largely data-driven techniques, while remaining sources of background are estimated from simulated MC event samples.
The shape and the normalisation of the $t\bar{t}\gamma$ background is cross-checked with a dedicated control region.

\subsection{\Zjets background}  \label{sec:zplusjetsbkg}
 
The background contribution from \Zjets production is estimated using a two-dimensional sideband method~\cite{STDM-2014-09}
based on considering together the probability that a jet satisfies the photon identification criteria and the probability that a jet satisfies the photon isolation criteria.
The \lplmg signal region is supplemented by three control regions which are disjoint from each other and from the signal region,
and which are dominated by \Zjets production.
Contributions to the control regions from \zg signal events and from non-(\Zjets) background are subtracted using estimates obtained from the MC event samples described in \Sect{\ref{sec:samples}}.
The fraction of \Zjets background events relative to the number of \zg signal events in the signal region
can be derived from the number of observed events in the signal and control regions according to the methodology described in \Ref{\cite{STDM-2014-09}}.
The relative fraction of \Zjets events is assumed to be the same for the \epemg and \mpmmg channels, and is determined by combining the two channels.
As a cross-check, the \Zjets fraction is determined separately for each channel, and the separate fractions are found to be consistent with each other.
In the case of differential cross-section measurements, the method is applied separately within each bin of the relevant kinematic observable,
giving a data-driven estimate of the shape as well as the rate of the \Zjets background.
 
The control regions are defined by modifying either the photon isolation requirements, or the photon identification requirements, or both.
Events in the signal region require the photon to satisfy FixedCutLoose isolation and Tight identification requirements, as described in \Sect{\ref{sec:objsel}}.
The modified photon identification criteria require that photon candidates fail to meet the Tight identification requirements
but satisfy {\textit{nontight}} selection criteria which remove requirements on
four\footnote{The four variables are $w_{{\mathrm{s}}3}$, $f_{\mathrm{side}}$, $\Delta E_{{\mathrm{s}}}$ and $E_{{\mathrm{ratio}}}$; their definitions are given in \Ref{\cite{PERF-2017-02}}.}
of the nine ECAL shower shape variables required for Tight photons.
The variables that are removed from the list of requirements are those that are least correlated with calorimeter isolation~\cite{STDM-2010-08}.
The modified photon isolation criteria select photon candidates that fail to satisfy the calorimeter-based component of the FixedCutLoose isolation requirements,
by requiring that \etiso is greater than $0.065\times\etg + \egap$, where $\egap$ is an offset separating the signal and non-isolated control regions, and is set to $2\,\GeV$.
The track-based component of the FixedCutLoose photon isolation requirements, $\ptiso < 0.05\times\etg$, is applied in all three control regions
(as well as in the signal region).
 
The contribution to each control region from \zg signal events is accounted for by using the \SHERPA MC signal sample to estimate
the fraction of signal events in the control region relative to the signal region.
These {\textit{signal leakage fractions}} are estimated to be approximately 6\% (1.5\%) for the control region with modified identification (isolation) criteria, and less than 0.2\% for the control region for which both the identification and isolation criteria are modified.
The contributions from non-(\Zjets) background to the signal and control regions are estimated from simulated MC samples, as described in \Sect{\ref{sec:otherbkg}}.
The non-(\Zjets) background fraction is estimated to be approximately 5\% for the signal region, and less than 2\% for each of the control regions.

The correlation between the probability that a jet satisfies the photon identification criteria and the probability that it satisfies the photon isolation criteria
is obtained from simulation using the \POWHEG MC \Zjets sample described in \Sect{\ref{sec:samples}}.
The fraction of \Zjets events satisfying the photon isolation requirement $\etiso < 0.065\times\etg$ in simulation
is greater for events satisfying the Tight photon identification criteria than for those failing to satisfy the Tight but satisfying the nontight criteria,
by a factor $R = 1.33 \pm 0.06$, where the uncertainty is the statistical uncertainty due to the limited number of MC events.
A value $R = 1$ would correspond to there being no correlation between the probabilities that a jet satisfies the photon identification criteria and the photon isolation criteria.
Systematic uncertainties in the ratio $R$ are studied by comparing data with simulation for events which satisfy the requirements defining the signal and control regions,
except that they fail to satisfy the track-based photon isolation requirement $\ptiso < 0.05\times\etg$,
resulting in event samples dominated by \Zjets events in all regions.
The ratio $R$ measured in data using these events, $R = 1.28 \pm 0.05$, is found to agree with the ratio predicted using the \POWHEG \Zjets MC sample, $R = 1.21 \pm 0.03$,
where in both cases the error is the statistical uncertainty.
The difference between these values is assigned as a systematic uncertainty in the ratio $R$, giving a total uncertainty in $R$ of $\pm 0.09$.
The value of $R$ determined above is significantly greater than unity, indicating a correlation between the photon identification and isolation criteria for jets.
This is found to be a result of the implementation of \etg-dependent Tight photon identification criteria for the analysis of \RunTwo data, as described in \Ref{\cite{EGAM-2018-01}},
together with the effect of the SR selection requirement on \etg.
 
Additional sources of systematic uncertainty in the \Zjets background estimate arise from uncertainties in the non-(\Zjets) background subtraction,
from uncertainties in the signal leakage fractions due to imperfect modelling of photon identification and isolation,
and from statistical uncertainties associated with the finite size of the MC sample used to determine the signal leakage fractions.
The overall relative uncertainty in the estimated \Zjets background is 11\%, of which the largest contribution (7\%) is due to the correlation uncertainty.
Cross-checks of the assigned uncertainty are carried out by varying the parameter \egap\ to 1\,\GeV\ and 3\,\GeV,
and by varying the number of ECAL shower shape variables which are removed in defining the nontight photon identification.
No additional uncertainty was found to be required as a result of these studies.
 
The background estimation presented above yields the event count $N_{\Zjets}$, which includes all \Zjets background, regardless of whether the jet identified as a photon comes from the hard scattering or from an additional pile-up interaction. The part of this background from pile-up jets is addressed in more detail in the following section.
\subsection{Pile-up background}  \label{sec:PUbkg}
 
Whereas the charged-particle tracks corresponding to the selected lepton pair are required to originate from the primary vertex,
no explicit requirement is imposed on the point of origin of the selected photon,
as this is, in general, relatively poorly measured, with an uncertainty which is much greater than the average spacing between the primary vertex candidates in the event.
This results in a small, but non-negligible, pile-up background
where a lepton pair produced in the \pp interaction giving rise to the primary vertex
combines with a photon produced in a second, independent, \pp interaction occurring in the same LHC bunch crossing.
Pile-up photon background from out of time bunch crossing is negligible after the requirements applied to the photon candidates.
 
A new method, developed for this analysis, is used to estimate this background source based on the fact that for photons from pile-up interactions there is no correlation between the $z$-positions of the interactions producing the $Z$-boson and the photon, while for the hard-scatter interactions they are the same. A complication in the method arises from the fact that selected photons from pile-up interactions can also come from misidentified jets, as discussed in
\Sect{\ref{sec:zplusjetsbkg}}, and care must be taken not to double-count this component.
 
The fractional pile-up photon background contribution is defined as
\begin{equation}
\fpug=\frac{N_{\textrm{PU},\gamma}}{\Nobs} \,,
\end{equation}
where $N_{\mathrm{PU,\gamma}}$ is the number of events from pile-up interactions with a genuine prompt photon, and \Nobs is the observed number of events.
 
In the data, first the total fraction of selected pile-up photons, $\fpu$, is estimated, including both photons from hard scatter interactions and jets misidentified as photons,
\begin{equation}
\fpu=\frac{N_{\textrm{PU},\gamma}+N_\textrm{PU,jets}}{\Nobs}=\frac{\fpug}{1-\fjet} \,.
\label{eq:fpu}
\end{equation}
Here $N_{\mathrm{PU,jets}}$ is the number of pile-up background events coming from misidentified jets, and $\fjet=\frac{N_\textrm{PU,jets}}{N_{\textrm{PU},\gamma}+N_\textrm{PU,jets}}$ is the fraction of the pile-up background events that come from misidentified jets.
 
The fraction \fpu is estimated by considering the distribution in data
of the longitudinal separation $\Dz = z_\gamma - \zvtx$ between the reconstructed primary vertex position, \zvtx,
and the position, $z_\gamma$, of the reconstructed photon after extrapolation to the beam-axis using the reconstructed photon direction.
Events where the selected lepton pair and the selected photon arise from separate \pp interactions ({\textit{pile-up events}})
are expected to have a broader \Dz distribution than events due to \zg signal production,
or to background processes associated with a single \pp interaction ({\textit{single-\pp events}}).
The pile-up background estimation uses SR events containing converted photons where both tracks from the conversion vertex are reconstructed in the ID
and where the conversion point is measured to be within the volume of the silicon pixel detector,
by requiring that the reconstructed radial coordinate of the conversion vertex is less than 125\,mm ({\textit{pixel conversions}}).
For these photons, the longitudinal position $z_\gamma$ is especially well reconstructed
(the uncertainty in $z_\gamma$ is always less than 1\,mm, and typically less than 0.2\,mm)
and the photon $z_\gamma$ resolution has a relatively small impact on the reconstructed \Dz distribution.
The \Dz distribution for pixel conversion events selected in the SR in data is shown in \Fig{\ref{fig:pu}}.
 
A sample enhanced in pile-up interactions is obtained by selecting pixel conversion events with $|\Dz|>50$\,mm.
The shape of the \Dz distribution for the pile-up component is obtained by assuming that the distributions of $z_\gamma$ and \zvtx are identical and uncorrelated, taking both from the \zvtx distribution observed in data.
The \zvtx distribution for selected events in the SR is well described by a Gaussian distribution of width $\sigma(\zvtx)=35.5 \pm 0.2$\,mm,
where the uncertainty is the statistical uncertainty from a fit to the data, and the observed width reflects the longitudinal spread of the proton bunches in the LHC.
Since $\Dz=z_\gamma-\zvtx$, and both $\zvtx$ and $z_\gamma$ follow a Gaussian distribution with width $\sigma(\zvtx)$ and are uncorrelated for pile-up, the $\Dz$ distribution for pile-up
is expected to follow a Gaussian distribution with $\sigma(\Dz)=\sqrt{2}\times 35.5=50.2$~mm.
Correspondingly, the probability that $|\Dz|>50$~mm for pile-up events is estimated as $\PhighDzPU=32\%$.
Using this information, the number of pile-up events in the pixel conversion sample can be estimated:
\begin{equation}
N_\textrm{PU,pix-conv}=\frac{\NhighDzdata - \NhighDzsinglepp}{\PhighDzPU} \,,
\end{equation}
where $\NhighDzdata=219$ is the number of data events with $|\Dz|>50$~mm ($\mathrm{high}\,|\Delta z|$) in the pixel conversion sample.
 
The term $\NhighDzsinglepp$ accounts for events from a single $pp$ interaction that pass the $\mathrm{high}\,|\Delta z|$ requirement.
It is estimated using the \SHERPA $Z\gamma$ MC sample, but rescaled by a correction factor derived in a control sample of $Z\to \ell\ell\gamma$ events,
selected by requiring  $86<\mllg<96$\,\GeV, instead of $\mll+\mllg>182$\,\GeV, to account for the somewhat wider $\Dz$ distribution in data compared to simulation.
In order to increase the statistical precision of this correction, the requirement on $\etg$ is relaxed to $\etg>15$~GeV.
The \Dz distribution for pixel conversion events in the $\Zboson\to\ell\ell\gamma$ control sample is shown in \Fig{\ref{fig:pu}}.
In this event sample, the contamination from pile-up background is expected to be negligible.
The number $\NhighDzsinglepp$ is determined to be $65 \pm 14$~events, where the uncertainty is dominated by the finite statistical precision of the control region.
To obtain $\fpu$, $N_\textrm{PU,pix-conv}$ needs to be divided by the total number of events ($10491$) with pixel conversion photons, resulting in $\fpu=(4.6\pm 0.6)\%$.
 
\begin{figure}
\begin{center}
\includegraphics[width=0.48\textwidth]{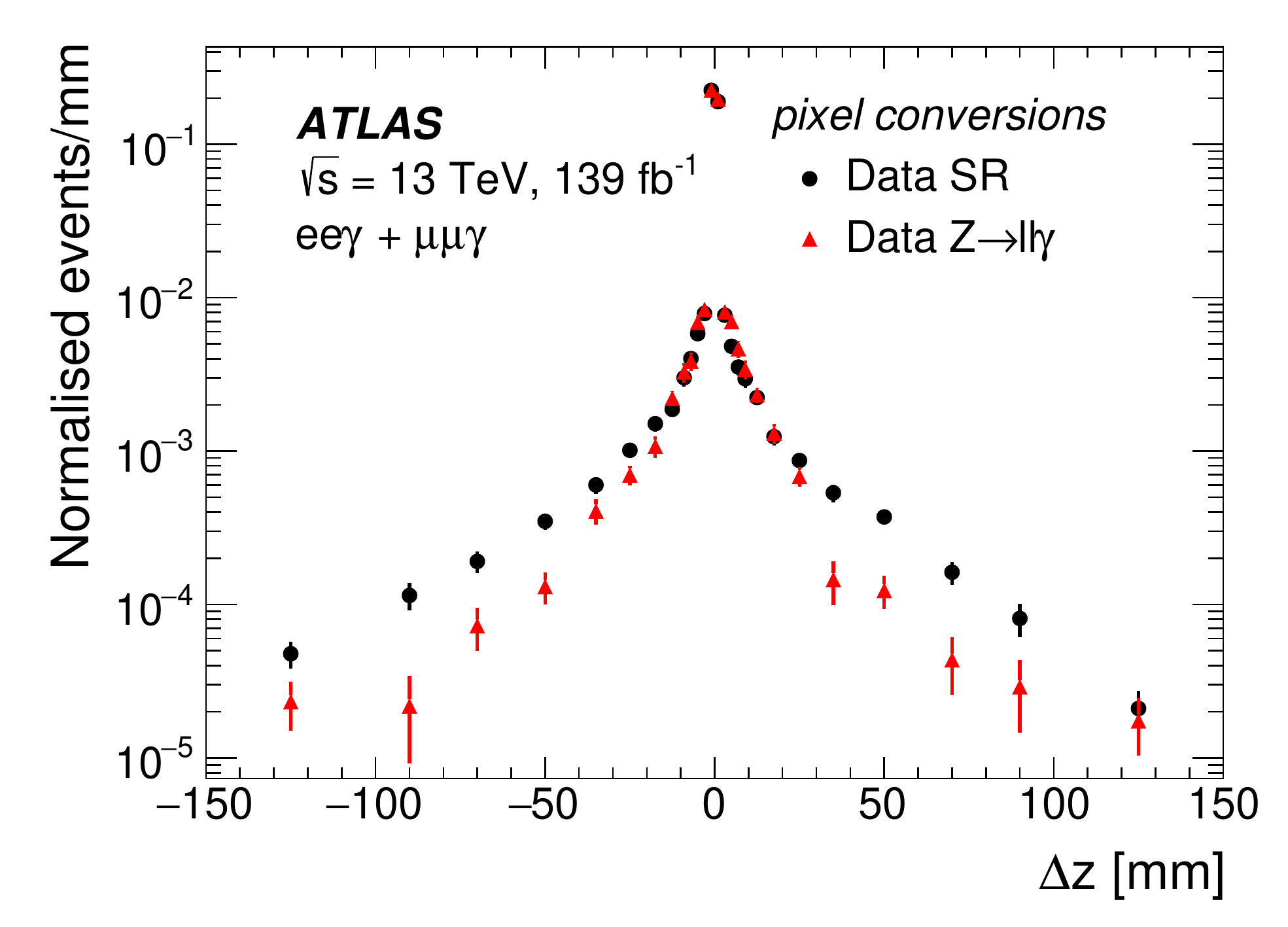}
\includegraphics[width=0.48\textwidth]{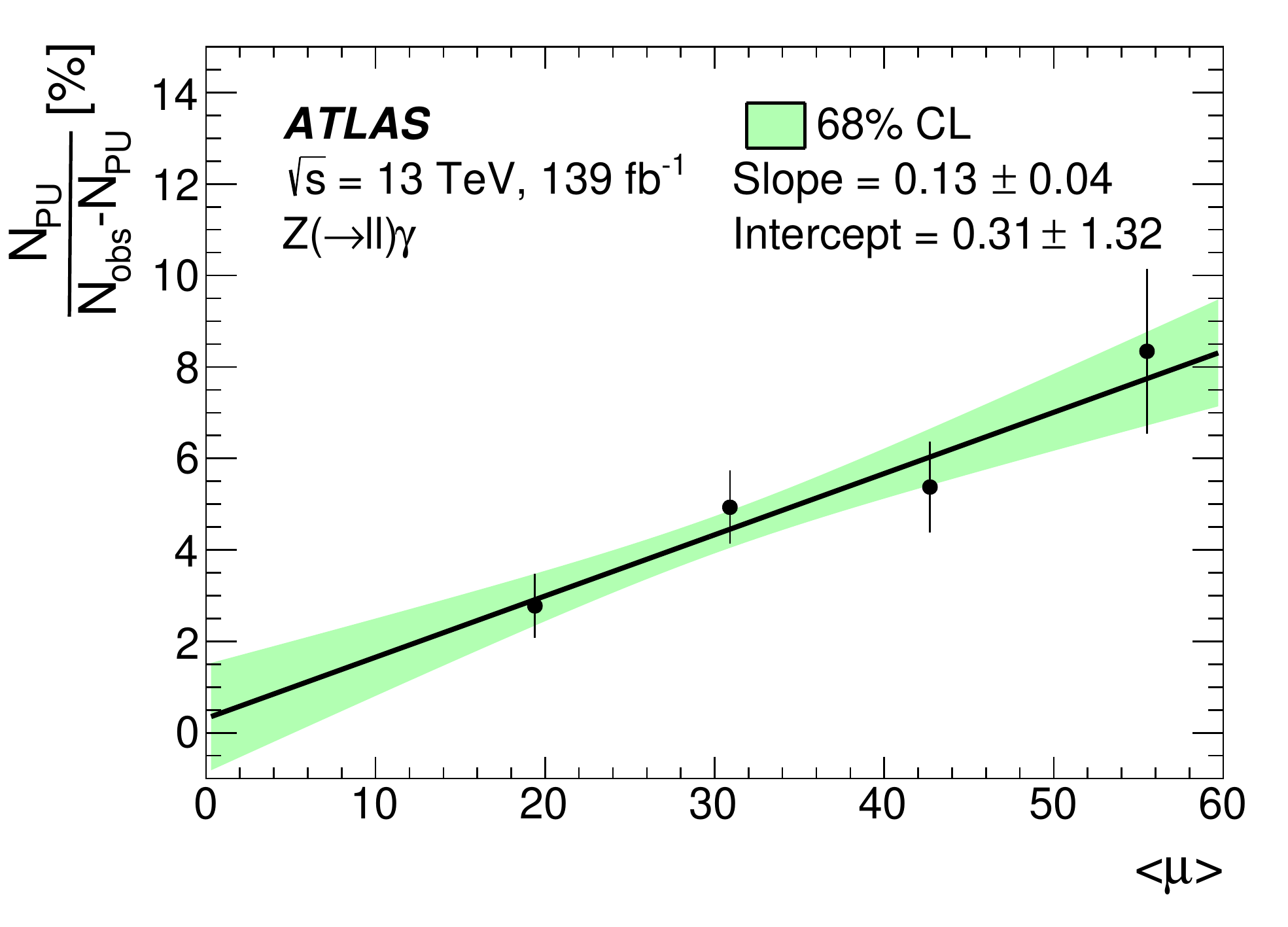}  \\
\caption{Left: Distributions of $\Dz$ for pixel conversion photons in the SR and in the $Z\to\ell\ell\gamma$ control region. Right: The ratio of the number of events where the photon candidate arises from a pile-up interaction to that where it arises from the same interaction as the $Z$ boson, is shown versus \meanmu. A straight-line fit to the data is also shown, and the intercept and the slope of the fit are given in the figure. The error bars on the ratios are uncorrelated between different values of \meanmu, and are due to the limited number of data and MC events. The shaded band shows the effect of the uncertainties in the fitted parameters.
}
\label{fig:pu}
\end{center}
 
\end{figure}
 
As stated above, this estimate contains both photons and misidentified jets, and needs to be corrected by a factor of $(1-\fjet)$, according to \Eqn{\ref{eq:fpu}}. Since the main source of isolated photons in these pile-up interactions is inclusive single-photon production occurring in the same bunch crossing as an inclusive $Z$ boson production event, this factor is determined in an inclusive sample of pixel conversion photons in data, using the two-dimensional sideband method introduced in \Sect{\ref{sec:zplusjetsbkg}}. Using this method, the fraction of events due to misidentified jets is estimated to be $\fjet = (46 \pm 7)$\%, where the uncertainty is the combined statistical and systematic uncertainty.
 
Finally, $\fpug=\fpu(1-\fjet)$ can be calculated, and is found to be $\fpug=(2.5\pm 0.5)$\%.
This is the measured fraction of pile-up photon events in the sample of SR events containing a pixel conversion.
Assuming that the fraction of events containing a pixel conversion is the same for pile-up photon and single-\pp interactions,
the fraction \fpug is also applicable to the entire sample of SR events.
The probability that a photon converts in the pixel detector and is reconstructed as a pixel conversion
is expected to be approximately independent of whether the photon is produced in the primary or a pile-up interaction.
However, the reconstruction efficiency for conversions is weakly dependent on the photon energy~\cite{EGAM-2018-01},
and differences between the prompt photon energy spectra for pile-up and single-\pp processes
could result in a difference between the corresponding fractions of pixel conversion events.
From a comparison of the pixel conversion fractions in simulated samples of inclusive photon and \zg signal events,
the uncertainty in \fpug for the full SR sample due to such an effect is found to be negligible in comparison to other sources of systematic uncertainty.
The number of pile-up background events in the SR from prompt photons is then obtained as $N_{\mathrm{PU,\gamma}}=\fpug\times\Nobs$,
and is given in \Tab{\ref{tab:backgroundsummary}}.
The estimated number of pile-up background events from misidentified jets, $N_{\mathrm{PU,jets}}$, is not required directly
as it is already part of the $N_{\Zjets}$ estimate described in the previous section.
It can nevertheless be calculated from $N_{\mathrm{PU,jets}}=(\fpu-\fpug)\times\Nobs$, and amounts to about 20\% of the $N_{\Zjets}$ background in both channels.
It is also given in \Tab{\ref{tab:backgroundsummary}}.
 
Cross-checks of the pile-up background estimation are carried out by varying the requirement on $|\Dz|$ used to define the pile-up--enhanced region within the range 25--100\,mm,
by using selected photons which are not pixel conversions but which have an uncertainty in the reconstructed position $z_\gamma$ less than 2\,mm,
and by estimating \fpug for the electron and muon channels separately.
No additional systematic uncertainty in \fpug is found to be required as a result of these cross-checks.
In addition, the ratio of the number of events with photon candidates (both prompt photons and fake photons) originating from pile-up interactions to that from single $pp$ interactions
is determined in four bins of \meanmu, as shown in \Fig{\ref{fig:pu}}. A fit to a straight line models the data well, and gives an intercept consistent with zero, as one would expect for pile-up.
 
An independent estimate of \fpug is obtained by taking the pile-up cross-section, \sigpu, to be given by $\sigpu = \meanmu \sigma_Z\sigma_\gamma/\sigin$,
where $\sigma_Z$ ($\sigma_\gamma$) is the cross-section for the inclusive production in \pp collisions
of a \Zboson boson (photon) satisfying the kinematic constraints summarised in \Tab{\ref{table:ZgSigReg}},
and $\sigin \approx 80$\,mb is the cross-section for inelastic \pp collisions.
The efficiency for pile-up events to satisfy the SR selection requirements is estimated from the \SHERPA LO \zg signal MC sample,
with the \etg spectrum reweighted to match that observed in the single-photon data sample.
This gives an estimate of \fpug consistent with that obtained from the \Dz distribution, within a relative uncertainty of about 30\%.
 
For the differential cross-section measurements,
the shapes of the relevant reconstructed kinematic distributions for pile-up background events
are estimated from a sample of simulated pile-up events, where each event is obtained by merging, at particle level,
the lepton pair from an event in the \Zjets \POWHEG sample with the prompt photon from an event in an inclusive photon sample generated using \SHERPAV{2.2.2} at NLO accuracy.
The kinematic requirements on the photon and the lepton pair summarised in \Tab{\ref{table:ZgSigReg}} are imposed on the merged event at particle level,
and bin-by-bin correction factors are applied to the particle-level distributions to model the effects of detector resolution and efficiency.
 
A related potential source of background arises from double-parton scattering (DPS),
in which the lepton pair and the photon are produced in separate parton--parton interactions occurring within the same \pp interaction.
The DPS cross-section, $\sigma_{\mathrm{DPS}}$, is estimated as $\sigma_{\mathrm{DPS}} \sim \sigma_Z\sigma_\gamma/\sigma_{\mathrm{eff}}$
where $\sigma_{\mathrm{eff}} \sim 15$\,mb is an empirical effective cross-section
(see \Ref{\cite{STDM-2017-15}}, for example).
This results in an estimated DPS background contribution of about 50~events per channel, which is at the per-mille level and  neglected.
 
 
\subsection{Other backgrounds}  \label{sec:otherbkg}
 
Background contributions from events due to \ttg, \zttg and $\Wboson\Wboson\gamma$ production, containing a genuine prompt photon,
and from $\Wboson\Zboson\to\ell\ell\ell\nu$ and $\Zboson\Zboson\to\ell\ell\ell\ell$ production, where an electron is misidentified as a photon,
are estimated using the simulated MC samples described in \Sect{\ref{sec:samples}}.
The process $\pp\to t\bar{t}\gamma+X$ contributes about 23\% of the total background,
while $\Wboson\Zboson$ production contributes about 4\%, and all other backgrounds each contribute less than 2\%.
 
The background contribution to the \lplmg signal region from \ttg production is estimated using the \MGMCatNLO LO \ttg MC sample described in \Sect{\ref{sec:samples}}.
The \ttg contribution to the \lplmg signal region obtained using this sample is multiplied by a normalisation factor of 1.44,
and a relative uncertainty of 15\% is assigned to the resulting background estimate.
This factor and its associated uncertainty were
determined in connection with an analysis of \ttg production at $\sqrt{s}=13$\,\TeV\ by the ATLAS Collaboration~\cite{TOPQ-2017-14},
and normalises the LO prediction from the \MGMCatNLO MC sample to an NLO calculation provided by the authors of \Ref{\cite{Melnikov:2011ta}}
for the fiducial phase-space region used for the \ttg measurement in the dilepton channel.
For the remaining background contributions to the \lplmg signal region estimated from MC event samples,
no additional normalisation factors are applied, and an uncertainty of 30\% is assigned to each estimated contribution.
This accounts for uncertainties in the inclusive cross-sections due to possible higher-order contributions,
and for experimental uncertainties such as those due to imperfect modelling of the probability that an electron is misidentified as a photon.
 
A small expected contribution (approximately 12 \epemg events and 15 \mpmmg events) from interactions containing a decay $H\to\zg$ of the Higgs boson is neglected.
 
\begin{figure}[t]
\begin{center}
\includegraphics[width=.495\columnwidth]{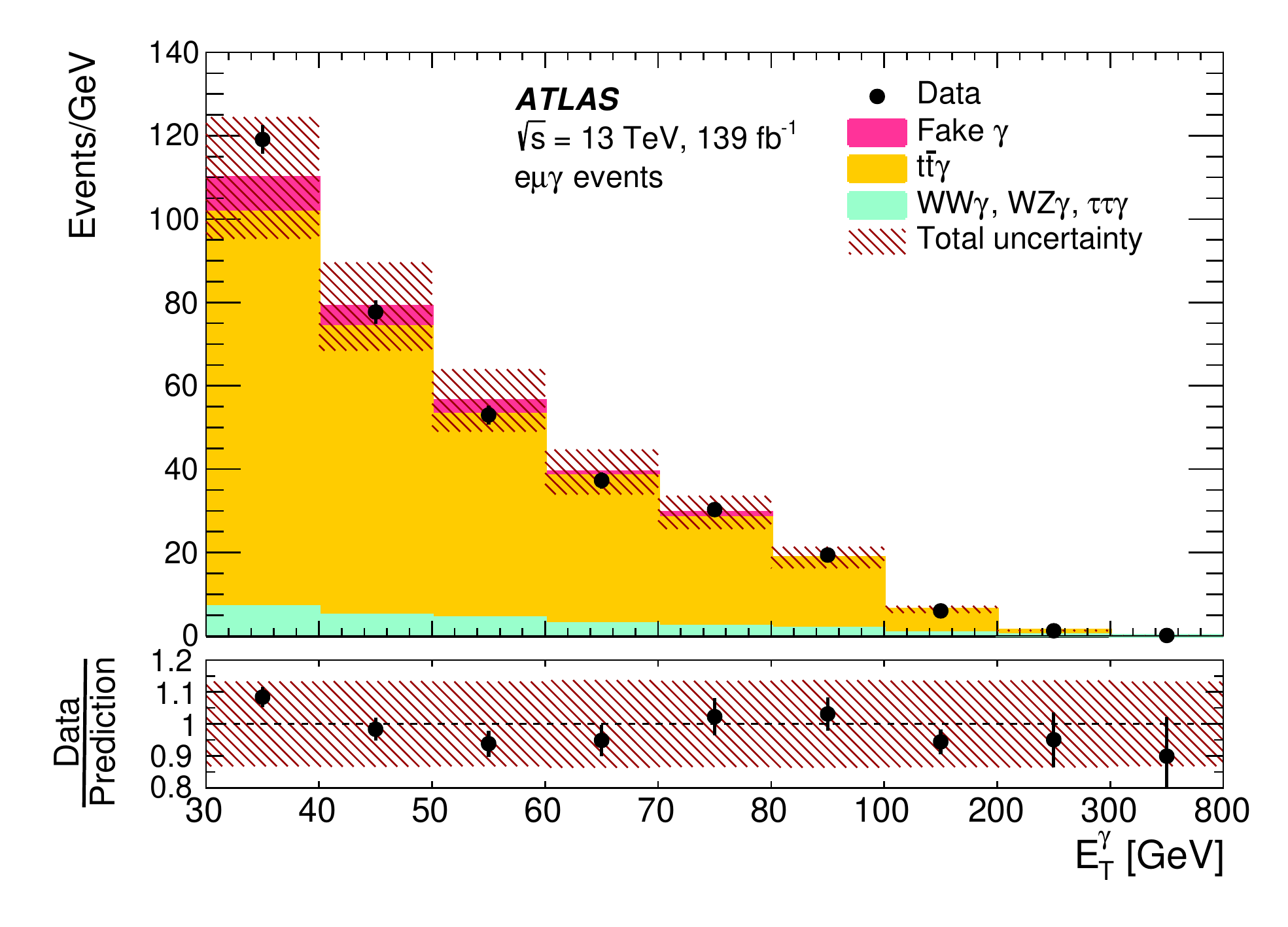}
\includegraphics[width=.495\columnwidth]{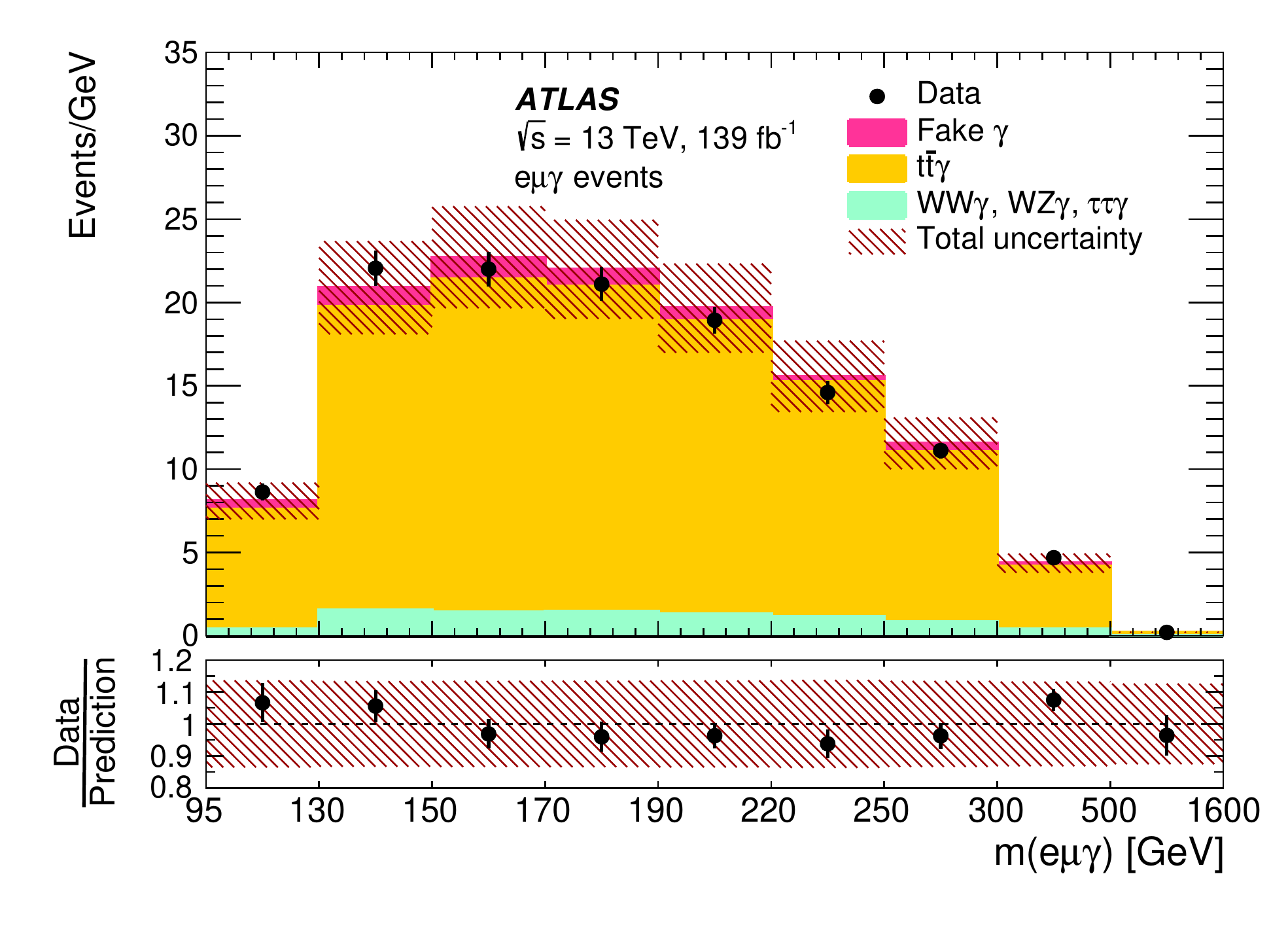}
\caption{Distributions of (left) \etg and (right) $m(e\mu\gamma)$ for selected \emg events.
The number of candidates observed in data (black data points) is compared with the sum of the expectation from \ttg, \wwg, \wzg, $\tau^+\tau^-\gamma$ and fake-photon background.
The lower panel in each plot shows the ratio of the observed and expected distributions.
The error bars on the observed distribution, and on the ratio of the observed distribution to the expected distribution, show the statistical uncertainty due to the number of observed events.
The hatched bands represent the total uncertainty on the expected distribution.
}
\label{fig:ttg}
\end{center}
\end{figure}
 
As a cross-check of the background estimation, a sample of opposite-charge, unlike-flavour \emg events is selected in data,
and compared with the expectation from the simulated MC background samples.
The contribution to the \emg sample from events in which a jet is misidentified as a photon ({\textit{fake-photon background}}) is also considered,
using a two-dimensional sideband method similar to that used above to estimate the \Zjets background contribution to the \epemg and \mpmmg signal samples.
The \emg sample is dominated ($\sim$90\%) by events due to \ttg production, while fake-photon background is estimated to contribute $\sim$4\% of the selected events.
A total of 4338 \emg events are selected, in agreement with a total background expectation of $4330 \pm 580$ events,
where the error is the combined statistical and systematic uncertainty.
The distributions of \etg and of the invariant mass, $m(e\mu\gamma)$, of the \emg system, are shown in \Fig{\ref{fig:ttg}},
and are observed to be in agreement with expectation within the total uncertainty in the expected number of events,
including the normalisation uncertainty of 15\% assigned to the predicted \ttg distributions.
 
\begin{table}[t]
\caption{
Summary of the observed number of events (\Nobs),
and the estimated number of background events ($N_{\Zjets}, N_{\mathrm{PU,\gamma}}, N_{\ttg}, N_{WZ},N_{ZZ}, N_{WW\gamma}, N_{\tau\tau\gamma}$),
in the \epemg and \mpmmg signal regions.
The $N_{\Zjets}$ background estimate includes a contribution from jets from pile-up interactions, $N_{\mathrm{PU,jets}}$, which is also shown separately.
In all cases, the uncertainty is the combination of the statistical and systematic uncertainties.
The bottom row gives the number of observed events after subtracting the sum, \Nbkg, of all estimated background contributions.
}
\label{tab:backgroundsummary}
\begin{center}
\begin{tabular*}{0.63\textwidth}
{l@{\extracolsep{\fill}}r@{ $\pm$\extracolsep{0.01cm} }r@{\extracolsep{\fill}}r@{ $\pm$\extracolsep{0.01cm} }r}
\hline
& \multicolumn{2}{c}{$e^+e^-\gamma$} & \multicolumn{2}{c}{$\mu^+\mu^-\gamma$}  \\
\hline
\Nobs                     &\multicolumn{2}{l}{ \hspace*{-3mm} 41343} &\multicolumn{2}{l}{\hspace*{-3mm} 54413} \\
\hline
$N_{\Zjets}$              &  4130 &  440     &     5470 & 580\phantom{)} \\
\hspace*{3mm}(includes $N_{\mathrm{PU,jets}}$     &  870 & 170      &     1140 &  230) \\
$N_{{\mathrm{PU}},\gamma}$  &  1030 & 210      &     1360 & 270\phantom{)}  \\
$N_{t\bar{t}\gamma}$          &  1650 &  250     &     1980 & 300\phantom{)}  \\
$N_{WZ}$                   &  254  &   76     &      199 & 60\phantom{)}  \\
$N_{ZZ}$                   &  64   &   19     &      102 & 31\phantom{)}  \\
$N_{WW\gamma}$              &  92   &   28     &      112 & 34\phantom{)}  \\
$N_{\tau\tau\gamma}$          &  46  &   15     &       39 & 12\phantom{)}  \\
\hline
$\Nobs - \Nbkg$           & 34080 &   590   &   45150 & 750\phantom{)}  \\
\hline
\end{tabular*}
\end{center}
\end{table}
 
 
\subsection{Background summary}  \label{sec:bkgsumm}
 
The estimated background yields in the \epemg and \mpmmg signal regions are summarised in \Tab{\ref{tab:backgroundsummary}}.

\begin{figure}[ht]
\begin{center}
\includegraphics[width=0.48\textwidth]{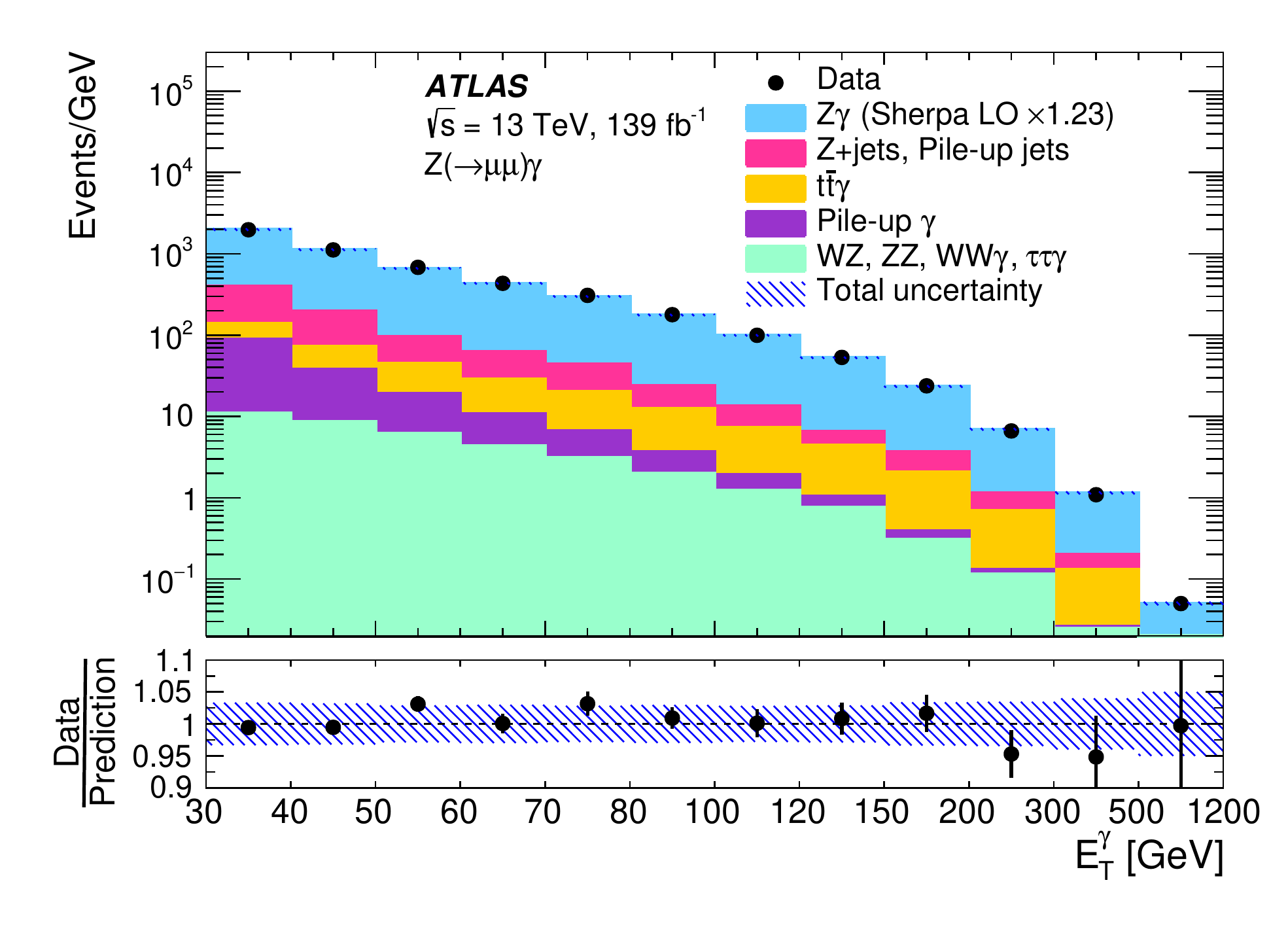}
\includegraphics[width=0.48\textwidth]{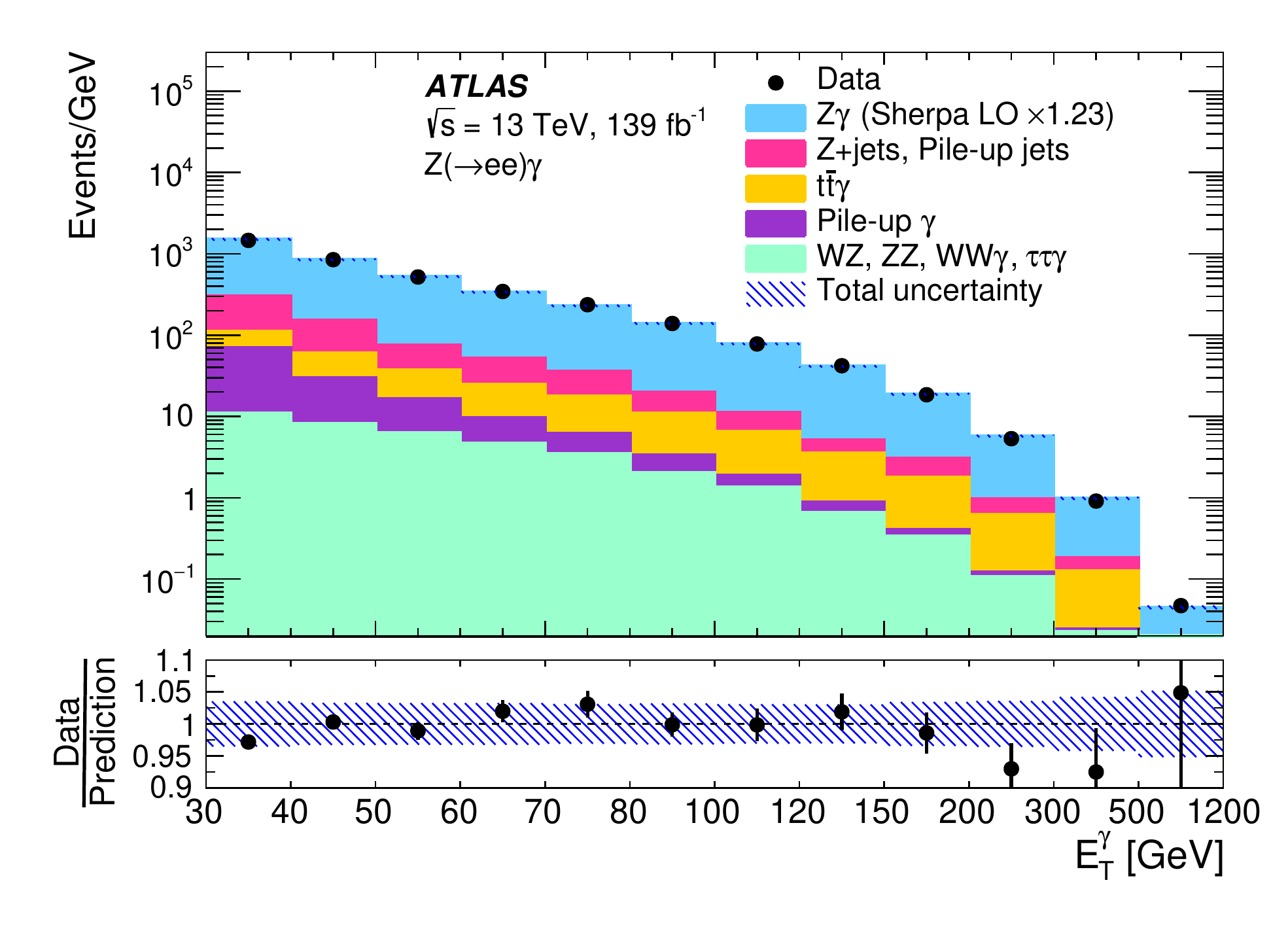}  \\
\includegraphics[width=0.48\textwidth]{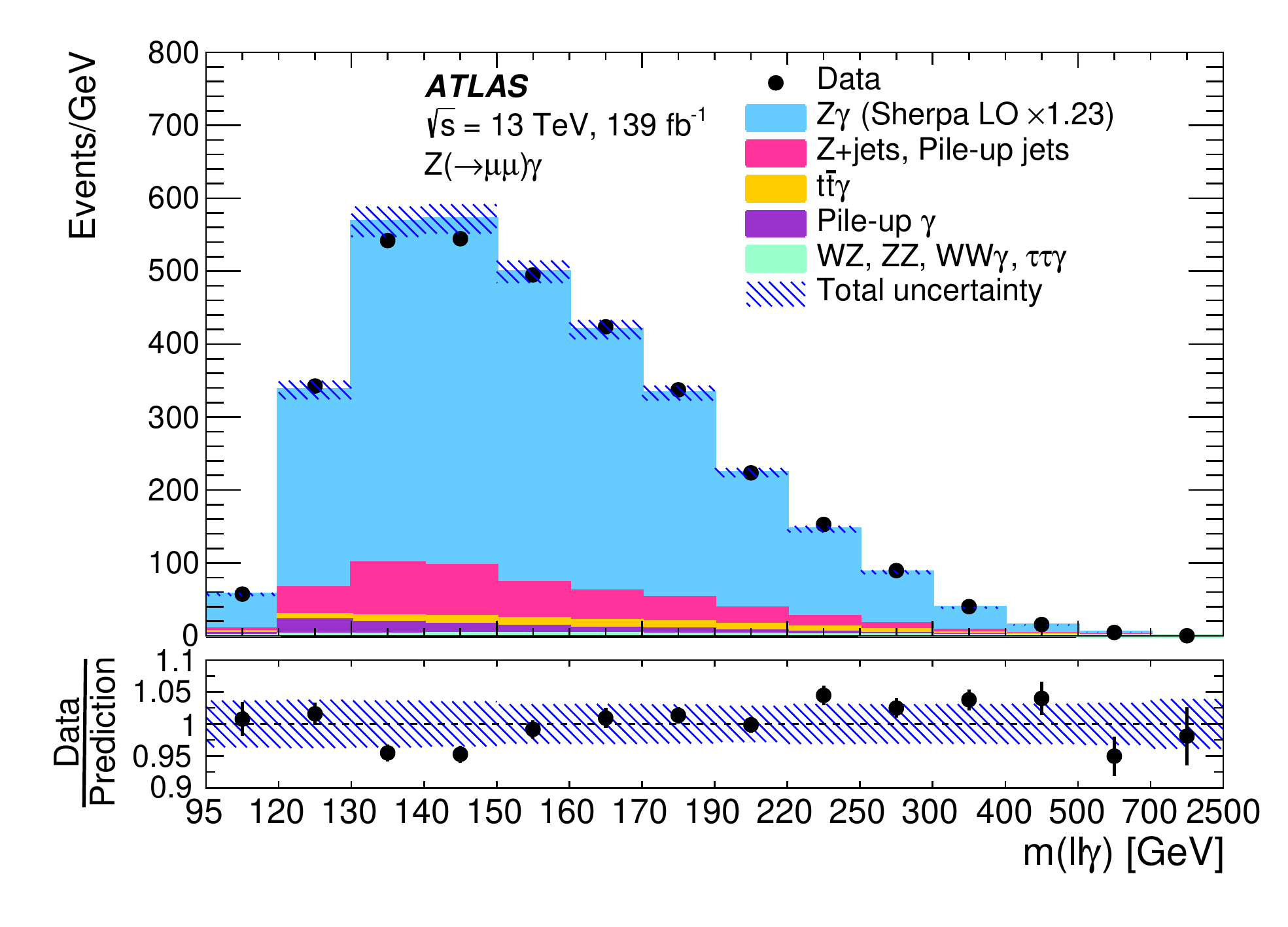}
\includegraphics[width=0.48\textwidth]{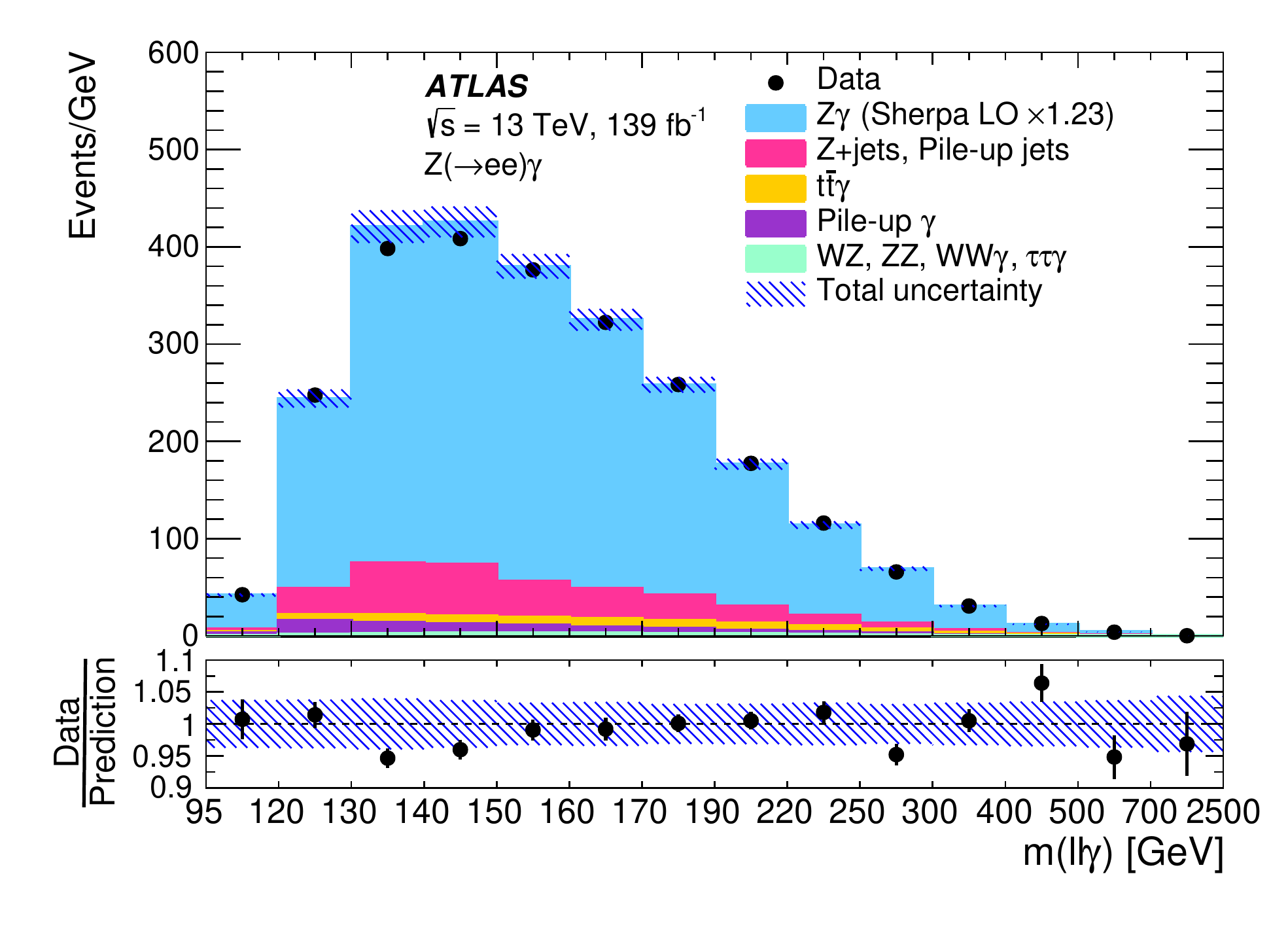}  \\
\caption{Distributions of (top) \etg and (bottom) \mllg for the (left) \mpmmg and (right) \epemg signal regions.
The number of candidates observed in data (black data points) is compared with
the sum of the signal predicted using the \SHERPA LO MC signal sample (including a normalisation factor of \SherpaLOnorm) and the estimated background contributions.
The lower section of each plot shows the ratio of the observed distribution to the sum of the predicted signal and estimated background.
The error bars on the observed distribution and on the ratio of the observed and expected distributions show the statistical uncertainty due to the number of observed events.
The hatched bands represent the sum in quadrature of the uncertainty in the background estimation, the statistical uncertainty in the MC signal prediction,
and the experimental systematic uncertainty, excluding the uncertainty in the integrated luminosity.
}
\label{fig:SR}
\end{center}
\end{figure}
 
\Fig{\ref{fig:SR}} shows the observed distributions of \etg and \mllg for events in the \epemg and \mpmmg signal regions,
together with the expected distributions for the \zg signal and for the background contributions.
A normalisation factor of \SherpaLOnorm is applied to the predicted contribution from the \SHERPA LO MC signal sample.
The normalisation factor is obtained from the ratio of the measured \lplmg cross-section to the cross-section predicted by \SHERPA at LO,
as presented in \Tab{\ref{table:ZgCrossSections}} in \Sect{\ref{sec:results_integxsec}}.
 
\section{Cross-section determination}
\label{sec:crosssection}
To simplify the interpretation of the results and the comparison with theoretical predictions, the \lplmg cross-section is measured
in a fiducial phase-space region defined by particle-level requirements similar to those defining the SR at reconstruction level,
and common to the \epemg and \mpmmg channels.
The requirements defining the fiducial region are summarised in \Tab{\ref{table:ZgFidReg}}.
Particle-level quantities are defined in terms of stable particles in the MC event record with a proper decay length $c\tau > 10$~mm which are produced from the hard scattering,
including those that are the products of hadronisation.
Compared to the SR,
the fiducial region imposes a common pseudorapidity selection ($|\eta| < 2.47$) on electrons and muons, and includes the ECAL barrel--endcap transition region in $|\eta|$ for photons and electrons.
For photons, the inclusion of the transition region
corresponds to a small interpolation ($\sim$6\%) within a slowly varying distribution.
The photon, and the electrons or muons, forming the \lplmg system
must not be produced in the decay of a hadron or a $\tau$-lepton.
The electron and muon four-momenta are corrected by adding the four-momenta of prompt photons within a cone of size $\Delta R = 0.1$ around each electron or muon, a procedure known as `dressing'.
Photon isolation at particle level is imposed by requiring the scalar sum of the transverse energy of all stable particles (except neutrinos and muons)
within a cone of size $\Delta R = 0.2$ around the photon, \etcone, to be less than 7\% of $\ET^{\gamma}$.
This upper limit
corresponds to the value of the ratio $\etcone/\etg$ for which there is an equal probability for simulated signal events to satisfy, or not satisfy,
the FixedCutLoose photon isolation requirements described in \Sect{\ref{sec:objsel}}.
No requirements are imposed at particle level on the electron or muon isolation.
 
\begin{table}[t]
\caption{Definition of the $\lplmg$ particle-level fiducial phase-space region.
For the lepton \pT requirements, the first (second) number specifies the minimum allowed \pT of the lepton with the highest (second-highest) value of transverse momentum.
}
\label{table:ZgFidReg}
\begin{center}
\begin{tabular}{ccc}
\hline
Photons  & & Electrons/Muons  \\
\hline
$\etg > 30$~\GeV                & & $\ptl > 30, 25$~\GeV  \\
$\aetag < 2.37$                 & & $|\eta^\ell| < 2.47$  \\
$\etcone/\etg < 0.07$           & & dressed leptons  \\
$\Delta R(\ell,\gamma) > 0.4$   & &  ~~  \\
\hline
\multicolumn{3}{c}{Event selection}  \\
\hline
\multicolumn{3}{c}{$\mll > 40\,\GeV$}  \\
\multicolumn{3}{c}{$\mll + \mllg > 182\,\GeV$}  \\
\hline
\end{tabular}
\end{center}
\end{table}
 
Measurements are made of the integrated \zg production cross-section in the particle-level fiducial region, and of the differential cross-sections
for six observables characterising the kinematic properties of the photon and the \lplmg system:
\etg, \aetag, \mllg, \ptllg, \rllg, and \dphillg.
For the differential cross-section measurements,
to minimise the dependence on the modelling of each distribution in the MC simulation,
an unfolding method is chosen to correct for the effects of detector inefficiency and resolution, as described in \Sect{\ref{sec:diffxsec}}.
For the integrated cross-section measurement,
the selection efficiency is taken directly from the signal MC sample, as described in \Sect{\ref{sec:integratedxsec}}.
All uncertainties are propagated consistently in both cases,
and the value of the integrated cross-section obtained from each differential measurement is found to be consistent with the central, directly obtained, value.
 
For all observables considered, the measured production rates for the electron and muon channels are found to be consistent with each other within their uncorrelated uncertainties.
The differential and integrated cross-section measurements in the electron and muon channels are averaged using
a $\chi^2$ minimisation method~\cite{Glazov:2005rn, Aaron:2009bp} in which correlations between bins and between the two channels are taken into account.
For each source of uncertainty which contributes to the total $\chi^2$, a nuisance parameter is introduced.
Correlated uncertainties are treated by using a common nuisance parameter for the \epemg and \mpmmg channels.

\subsection{Integrated fiducial cross-section measurement}  \label{sec:integratedxsec}
 
The integrated cross-section in the fiducial phase-space region defined in \Tab{\ref{table:ZgFidReg}} is calculated as
\begin{linenomath}
\begin{equation*}
\sigfid = \frac{\Nobs - \Nbkg}{C \times \intlumi} \,,
\end{equation*}
\end{linenomath}
where \Nobs is the observed number of selected events in the data in the signal region, \Nbkg is the expected number of background events,
\intlumi is the integrated luminosity corresponding to the analysed dataset, and the factor $C$ corrects for detection efficiency and acceptance.
The value of the numerator $\Nobs - \Nbkg$ for each channel is given in
\Tab{\ref{tab:backgroundsummary}}.
The correction factor $C$ is determined using the \epemg and \mpmmg simulated signal MC event samples generated using \SHERPAV{2.2.4} at LO.
It is defined as the number of reconstructed signal events satisfying all selection criteria
divided by the number of events that, at particle level, meet the acceptance criteria of the fiducial region.
The values of the correction factors $C$ for each channel are obtained as
$\Ceeg = 0.462 \pm 0.007\,({\mathrm{uncorr}}) \pm 0.008\,({\mathrm{corr}})$ and $\Cmmg = 0.607 \pm 0.005\,({\mathrm{uncorr}}) \pm 0.009\,({\mathrm{corr}})$
where, in each case,
the first error is the component of the uncertainty which is uncorrelated between the two channels,
and the second is the correlated component of the systematic uncertainty.
The systematic uncertainties are determined using the procedures described in \Sect{\ref{sec:systematics}}.
 
Due to measurement resolution effects, events lying within (outside) the fiducial region at particle level can migrate to lie outside (within) the SR after event reconstruction.
Such migrations are implicitly corrected for using the efficiency factors \Ceeg and \Cmmg,
but this relies on the simulation accurately describing the distributions of the variables used to define the SR.
The largest migrations occur for \etg,
and their possible impact is assessed by reweighting the \etg spectrum in the signal MC event sample to agree with that observed in data.
The difference between the efficiency factors obtained using the original or reweighted spectrum is less than 0.1\%.

\subsection{Differential fiducial cross-section measurements}  \label{sec:diffxsec}
 
The differential cross-sections in the fiducial region for each of the six observables \etg, \aetag, \mllg, \ptllg, \rllg and \dphillg,
are extracted using the unfolding procedure described in \Ref{\cite{STDM-2012-07}} to correct for measurement inefficiencies and resolution effects.
The unfolding procedure employs an iterative Bayesian method~\cite{DAgostini:1994fjx} with two iterations.
For each distribution, events from the \SHERPA simulated signal MC sample are used to generate a response matrix
that accounts for bin-to-bin migration between the reconstruction-level and particle-level distributions.
 
The statistical uncertainties in the unfolded distributions are estimated using pseudo-experiments,
generated by fluctuating each bin of the observed spectrum according to a Poisson distribution with a mean value equal to the observed yield.
The shape uncertainties arising from the limited size of the signal MC sample are also obtained by generating pseudo-experiments.
The sources of systematic uncertainty are discussed in \Sect{\ref{sec:systematics}}, with their impact on the unfolded distribution assessed by varying the response matrix
for each of the systematic uncertainty sources by one standard deviation and combining the resulting differences from the nominal values in quadrature.
As a cross-check of the unfolding procedure, a data-driven closure test is performed by reweighting the shape of the particle-level distributions in simulated MC event samples
with a smooth function chosen such that the reconstruction-level distribution for the MC sample closely reproduces that observed in data after the reweighting.
No additional systematic uncertainty is found to be required as a result of this test.

\subsection{Systematic uncertainties}  \label{sec:systematics}
 
Systematic uncertainties in the measured cross-sections arise from uncertainties in the correction factor $C$ and the unfolding procedure, uncertainties in the estimated background, \Nbkg,
and uncertainties in the integrated luminosity, \intlumi.
The uncertainties in \Nbkg and \intlumi are discussed in \Sects{\ref{sec:backgrounds}}{\ref{sec:samples}}, respectively.
Systematic uncertainties affecting the factor $C$ and the unfolding include contributions arising from uncertainties in the efficiencies of the trigger, reconstruction, and particle identification and isolation,
and from uncertainties in the energy and momentum scales and resolutions of reconstructed photons, electrons and muons.
 
The performance of the electron and photon reconstruction, and the associated systematic uncertainties, are studied in \Ref{\cite{EGAM-2018-01}}.
For electrons, the reconstruction, identification and isolation efficiencies, and their uncertainties,
are measured by applying tag-and-probe methods to events containing \Zepem or \Jee decays. 
For photons, the corresponding efficiencies are measured using
samples of $\Zboson\to\lplmg$ $(\ell = e, \mu)$ and \Zepem decays,
and an inclusive photon sample collected using single-photon triggers. 
The energy scale and resolution for electrons and photons, and their uncertainties, are obtained from a sample of \Zepem events
and cross-checked with samples of \Jee and $\Zboson\to\lplmg$ decays. 
For muons, the efficiencies, and the momentum scale and resolution, and their uncertainties, are obtained using samples of \Zmpmm and \Jmm decays~\cite{PERF-2015-10}.

A comparison of data with simulation for events satisfying the signal region requirements of \Tab{\ref{table:ZgSigReg}}, but with the requirement $\mll+\mllg > 182\,\GeV$ removed,
indicates a possible mismodelling, at the level of 25\%, of the relative rate of events which satisfy, or do not satisfy, this requirement in the \SHERPA MC signal sample.
The effect of such a mismodelling was assessed by varying the rate of events in the \SHERPA sample that do not satisfy the requirement $\mll+\mllg > 182\,\GeV$ at particle level by 25\%.
The effect on the measured integrated and differential cross-sections in the fiducial region is negligible in comparison with other sources of systematic uncertainty.
 
The systematic uncertainties in the integrated cross-section in the fiducial region, \sigfid, are summarised in \Tab{\ref{table:Systematics}}.
For all differential cross-sections, the largest systematic uncertainty arises from the background estimation. 
 
\begin{table}
\caption{Relative uncertainties in the measured integrated cross-section, \sigfid, for \lplmg production within the fiducial phase-space region defined in \Tab{\ref{table:ZgFidReg}}.
The upper section of the table lists the individual sources of systematic uncertainty,
followed by the total systematic uncertainty obtained by combining the individual contributions in quadrature.
Only sources which contribute a relative uncertainty of at least 0.1\% are listed.
An entry "--" indicates that the uncertainty source is not applicable to the given channel or the relative uncertainty is less than 0.1\%.
The rightmost column indicates whether the uncertainties for each source are fully correlated (`yes'), partially correlated (`partial') or uncorrelated (`no')
between the \epemg and \mpmmg channels.
The penultimate row gives the statistical uncertainty due to the number of observed events in the signal region.
The bottom row gives the overall relative uncertainty obtained by combining the systematic and statistical uncertainties in quadrature.
}
\label{table:Systematics}
\begin{center}
\begin{tabular}{lccl}
\hline
Source                            &   \multicolumn{2}{c}{Uncertainty [\%]}  &  Correlation  \\
\hspace*{65mm}                    &  ~\epemg~  &  ~\mpmmg~  &  \\
\hline
Trigger efficiency                    &  --  &  0.2   &  no  \\
Photon identification efficiency      & \multicolumn{2}{c}{1.0}  &  yes  \\
Photon isolation efficiency           & \multicolumn{2}{c}{0.9}  &  yes  \\
Electron identification efficiency    & 1.4  &   --   &  no   \\
Electron reconstruction efficiency    & 0.3  &   --   &  no  \\
Electron--photon energy scale         & 0.9  &  0.6   &  partial  \\
Muon isolation efficiency             &  --  &  0.4   &  no  \\
Muon identification efficiency        &  --  &  0.7   &  no  \\
\Zjets background                     & \multicolumn{2}{c}{1.3}  &  yes  \\
Pile-up background                    & \multicolumn{2}{c}{0.6}  &  yes  \\
Other backgrounds                     & 0.8  &  0.7   &  partial  \\
Monte Carlo event statistics          & 0.4  &  0.4   &  no  \\
Integrated luminosity                 & \multicolumn{2}{c}{1.7}  &  yes  \\
\hline
Systematic uncertainty                & 3.2  &  2.9   &  \\
Statistical uncertainty               & 0.6  &  0.5   &  \\
\hline
Total uncertainty                     & 3.2  &  3.0   &  \\
\hline
\end{tabular}
\end{center}
\end{table}

 
\section{Standard Model calculations}
\label{sec:SMpredictions}

The cross-section for the \zg process has been computed at NNLO in pQCD~\cite{Grazzini:2015nwa,Campbell:2017aul}.
The measured integrated and differential cross-sections are compared with predictions from the parton-level generator \MATRIX~\cite{Grazzini:2017mhc}, corrected to particle level, at both NLO and NNLO. The measured cross-sections are also compared with SM expectations obtained using the parton shower MC generators \SHERPA and \MGMCatNLO.

The predictions from the \SHERPA event generator at LO and from the \MGMCatNLO generator at NLO
are obtained using particle-level events from the signal MC samples described in \Sect{\ref{sec:samples}}.
The predictions from \SHERPA at NLO are obtained using \SHERPAV{2.2.8}, configured according to the \MEPSNLO setup described in \Ref{\cite{Krause:2017nxq}}.
In this setup, up to three additional final-state partons are generated where up to one additional final-state parton is at NLO accuracy,
and the \NNPDFVV{3.0}{NNLO} PDF set is used.
For the predictions obtained using \SHERPA or \MGMCatNLO, only the statistical uncertainty due to the limited number of MC events generated is considered.
The predictions from \MATRIX are obtained for the CT14nnlo PDF set~\cite{Dulat:2015mca}, and using the transverse momentum ($q_\mathrm{T}$) subtraction method~\cite{Catani:2007vq}.
The values of the renormalisation and factorisation scales are set to $\sqrt{\mll^2 + (\etg)^2}$~\cite{Grazzini:2017mhc}.
For all predictions, smooth-cone photon isolation is imposed at parton level with the same choice of parameters ($\frixdelmax = 0.1$, $\frixepsmax = 0.1$, $n = 2$; see \Sect{\ref{sec:samples}})
as used in the generation of the \SHERPA LO MC signal sample.

Electroweak (EW) radiative corrections to \zg production have been computed at NLO~\cite{Denner:2015fca, Hollik:2004tm, Accomando:2005ra},
including for the fiducial phase-space region defined in \Tab{\ref{table:ZgFidReg}},
both inclusively and as a function of the observables \etg, \aetag and \mllg\cite{Denner:2015fca}.
The EW corrections are provided separately for partonic processes with a $q\overline{q}$, $q\gamma$ or $\gamma\gamma$ initial state.
Their impact on the NNLO cross-section predicted by \MATRIX is considered.
The absence of a complete, combined calculation of NLO EW and NNLO QCD corrections
results in an ambiguity as to whether the NLO EW corrections
associated with the $q\overline{q}$ initial state should be applied multiplicatively or additively to the NNLO QCD corrections computed using \MATRIX~\cite{Denner:2015fca}.
Both the multiplicative and additive approaches are considered in comparing the theoretical predictions with measurement.
 
The parton-level cross-section predictions from \MATRIX are corrected to particle level by applying parton-to-particle correction factors, \Cptop.
These correction factors are computed using parton-level and particle-level events from the \SHERPA LO signal MC sample described in \Sect{\ref{sec:samples}}.
The factor \Cptop is obtained as the ratio of the $\pp\to\lplmg$ cross-section predicted by \SHERPA
at particle level within the fiducial phase-space region defined in \Tab{\ref{table:ZgFidReg}}
to the predicted cross-section at parton level within a fiducial region defined as in \Tab{\ref{table:ZgFidReg}}
but with the smooth-cone isolation prescription defined above replacing the particle-level photon isolation criterion.
In the case that EW corrections are not applied to the \MATRIX prediction,
the parton-level fiducial region is defined using Born-level leptons in place of dressed leptons.
The systematic uncertainty in \Cptop is evaluated from a comparison with the correction factor obtained using events generated with \SHERPAV{2.2.2} at NLO.
The value of \Cptop\ obtained when EW corrections are applied (not applied) is $0.934 \pm 0.005$ ($0.915\pm 0.009$) for the integrated cross-section,
and varies between $0.83$ and $0.99$ ($0.76$ and $0.98$) across all bins used for the differential cross-section measurements.

For the predictions from \MATRIX at NLO and NNLO,
the uncertainties arising from the choice of PDF set and the value of \alphas are assessed according to the PDF4LHC recommendations~\cite{Butterworth:2015oua}.
The PDF uncertainty is evaluated using the PDF set NNPDF30\_nnlo\_as\_0118~\cite{Ball:2014uwa},
and the \alphas uncertainty is evaluated using the PDF sets NNPDF30\_nnlo\_as\_0117 and NNPDF30\_nnlo\_as\_0119.
The uncertainty associated with the choice of  renormalisation (\mur) and factorisation (\muf) scales is also considered.
The scale uncertainty is evaluated by varying \mur and \muf independently
by factors of 2 and 0.5 from their nominal values, with the constraint $0.5 \le \muf/\mur \le 2$.
The envelope of the resulting variations is taken as the size of the associated systematic uncertainty.
 
There is no accepted prescription for assigning a systematic uncertainty associated with the choice of photon isolation criteria imposed at parton level.
For illustrative purposes, for the smooth-cone prescription,
decreasing the value of the maximum cone size \frixdelmax from 0.1 to 0.05 increases the predicted fiducial cross-section by approximately 2.2\%,
while increasing the value of the parameter \frixepsmax from 0.1 to 0.2 leaves the predicted cross-section unchanged, within a statistical precision of 0.5\%.
The choice of parton-level photon isolation criteria used in the generation of the signal MC sample
potentially affects the estimated value of the correction factor $C$, and hence also the measured fiducial cross-section \sigfid.
Using an alternative \SHERPA LO MC signal sample generated with a smooth-cone isolation requirement which is much tighter ($\frixdelmax=0.3$, $\frixepsmax=0.025$, $n=2$)
than that used for the baseline sample is found to leave the correction factors \Ceeg\ and \Cmmg\ unchanged, within a statistical precision of 0.7\%.
 
A small expected SM contribution from the electroweak production of a \zg pair in association with two jets, $qq\to\zgjj$,
which includes the vector-boson scattering subprocess $\Wboson^+\Wboson^-\to\zg$, is also considered~\cite{STDM-2017-26, CMS-SMP-14-018}.
This contribution is evaluated at LO accuracy using the \MGMCatNLOV{2.3.3} generator with no extra parton in the final state, and interfaced to \PYTHIA for hadronisation.
The PDF set NNPDF30\_nlo\_as\_0118 is used, and the factorisation scale is set to the invariant mass of the diboson system.

\section{Results}
\label{sec:results}
\subsection{Integrated fiducial cross-section}   \label{sec:results_integxsec}
 
The measured cross-sections for \zg production in the fiducial phase-space region defined in \Tab{\ref{table:ZgFidReg}}
for the \epemg and \mpmmg channels
are given in \Tab{\ref{table:ZgCrossSections}}.
The uncertainties in the \epemg and \mpmmg cross-sections include components $\pm 9.0$\,fb and $\pm 6.1$\,fb, respectively, which are uncorrelated between the two channels.
The \epemg and \mpmmg cross-sections are consistent within the uncorrelated uncertainties, and are averaged using the procedure described in \Sect{\ref{sec:crosssection}}.
The resulting measured cross-section for \lplmg production is
 
\vspace*{-5mm}    
 
\[  \sigfid = 533.7 \pm 2.1 ({\mathrm{stat}}) \pm 12.4 ({\mathrm{syst}}) \pm 9.1 ({\mathrm{lumi}})\,\,{\mathrm{fb}}\,.  \]
The overall relative precision of the cross-section measurement is 2.9\%.
 
The measured cross-sections are compared with particle-level theoretical predictions obtained from the parton shower generators \SHERPA and \MGMCatNLO,
and from the parton-level generator \MATRIX corrected to particle level, as described in \Sect{\ref{sec:SMpredictions}}.
The predicted cross-sections are summarised in \Tab{\ref{table:ZgCrossSections}}.

\begin{table}[t]
\caption{Measured cross-sections (first three rows) for \lplmg production within the particle-level fiducial phase-space region defined in \Tab{\ref{table:ZgFidReg}},
compared with (next five rows) corresponding SM expectations obtained from the \SHERPA event generator at LO and NLO, \MGMCatNLO event generator at NLO,
and from the \MATRIX generator at NLO and NNLO.
For the measured cross-sections, the first uncertainty is due to all sources which are uncorrelated between the \epemg and \mpmmg channels (including the statistical uncertainty),
while the second is the remaining systematic uncertainty, excluding the uncertainty in the integrated luminosity, shown separately.
For the predicted cross-sections, the first uncertainty is due to the finite number of generated events,
the second is the uncertainty due to the correction factor \Cptop, the third is the uncertainty associated with the choice of PDF and the value of \alphas,
and the final uncertainty is due to the choice of renormalisation and factorisation scales.
The SM cross-section for EW \zgjj production
is included in all cross-section predictions. The NLO EW radiative corrections are applied to the \MATRIX NNLO cross-section multiplicatively and additively in the last two rows.
}
\label{table:ZgCrossSections}
\begin{center}
\begin{tabular}{lrll}
\hline
&  \multicolumn{3}{c}{Cross-section [fb]}  \\
\hline
$e^+e^-\gamma$             &  530.4 & $\pm$ 9.0 (uncorr)  & $\pm$ 11.7 (corr) $\pm$ 9.0 (lumi) \\
$\mu^+\mu^-\gamma$         &  535.0 & $\pm$ 6.1 (uncorr)  & $\pm$ 11.5 (corr) $\pm$ 9.1 (lumi) \\
\hline
$\ell^+\ell^-\gamma$       &  533.7 & $\pm$ 5.1 (uncorr)  & $\pm$ 11.6 (corr) $\pm$ 9.1 (lumi) \\
\hline
\SHERPA LO                 &  438.9 & $\pm$ 0.6 (stat)  &    \\
\SHERPA NLO                &  514.2 & $\pm$ 5.7 (stat)  &    \\
\MADGRAPH NLO              &  503.4 & $\pm$ 1.8 (stat)  &    \\
\MATRIX NLO                &  444.2 & $\pm$ 0.1 (stat) & $\pm$ 4.3 (\Cptop) $\pm$  ~\,8.8 (PDF) $^{+16.8}_{-18.9}$ (scale)  \\
\MATRIX NNLO               &  518.9 & $\pm$ 2.0 (stat) & $\pm$ 5.1 (\Cptop) $\pm$    10.8 (PDF) $^{+16.4}_{-14.9}$ (scale)  \\
\MATRIX NNLO $\times$ NLO EW & 513.5 & $\pm$ 2.0 (stat) & $\pm$ 2.7 (\Cptop) $\pm$    10.8 (PDF) $^{+16.4}_{-14.9}$ (scale)  \\
\MATRIX NNLO $+$ NLO EW      & 518.3 & $\pm$ 2.0 (stat) & $\pm$ 2.7 (\Cptop) $\pm$    10.8 (PDF) $^{+16.4}_{-14.9}$ (scale)  \\
 
\hline
\end{tabular}
\end{center}
\end{table}

The measured \lplmg cross-section is about 20\% higher than the predictions from \SHERPA at LO and from \MATRIX at NLO, about 6\% higher than the prediction from \MGMCatNLO and about 4\% higher than the prediction from \SHERPA at NLO.
The \MATRIX, \SHERPA and \MADGRAPH NLO predictions, although formally of the same order, cannot be compared directly
as the latter two are based on multi-leg MC event generators which include additional LO processes producing hard QCD radiation.
The measured cross-section is about 3\% higher than the prediction from \MATRIX at NNLO, and consistent with it within about $0.7\sigma$.
The correction to the predicted \MATRIX cross-section at NNLO compared to NLO is about $+17$\%, and is significantly larger than the scale uncertainty estimated at NLO.
Such an effect is discussed in Ref.~\cite{Grazzini:2015nwa},
where it is noted that, due to LO kinematic effects, the higher-order correction is enhanced by increasing the requirement on \etg.
 
\Tab{\ref{table:ZgCrossSections}} also gives the \MATRIX NNLO cross-sections as modified by the multiplicative and additive NLO EW corrections, as discussed in \Sect{\ref{sec:SMpredictions}}.
NLO EW radiative corrections are predicted to reduce the \MATRIX NNLO cross-section by as much as about $-1$\%, although with a large uncertainty,
as illustrated by the difference between applying the $q\overline{q}$ component of the EW corrections multiplicatively or additively, which produce shifts of $-8.2$~fb and $-3.4$~fb respectively, in addition to smaller shifts of $+2.5$~fb and $+0.3$~fb from $\gamma\gamma$- and $q\gamma$-induced production.
The cross-section for EW \zgjj production is predicted to be $4.57 \pm 0.02$\,fb,
where the uncertainty is due to the limited number of generated events.
The \zgjj contribution is included in all predicted cross-sections shown in \Tab{\ref{table:ZgCrossSections}}.

\subsection{Differential fiducial cross-sections}   \label{sec:results_diffxsec}
 
The measured and predicted differential cross-sections as a function of each of the quantities \etg, \aetag, \mllg, \ptllg, \rllg and \dphillg
are shown in \Figs{\ref{fig:UnfoldedDataVsSHERPA}}{\ref{fig:UnfoldedDataVsMATRIX}}.
The measurements cover an \etg range up to 1.2~\TeV\ and an \mllg range up to 2.5~\TeV.
The distribution of \ptllg has a maximum near 10~\GeV\ and falls slowly for higher \ptllg values.
The \dphillg distribution shows that, for the majority of events, the $Z$ boson and photon are produced approximately back-to-back,
but there are a significant number of events where they are close to each other in azimuth.
The relative precision of the differential cross-section measurements is in the range 3--7\% in all bins,
except for the highest two bins in \etg where, due to the limited number of events in data, it approaches about 15\%.
 
\begin{figure}[hbtp]
\begin{center}
\includegraphics[width=0.495\textwidth]{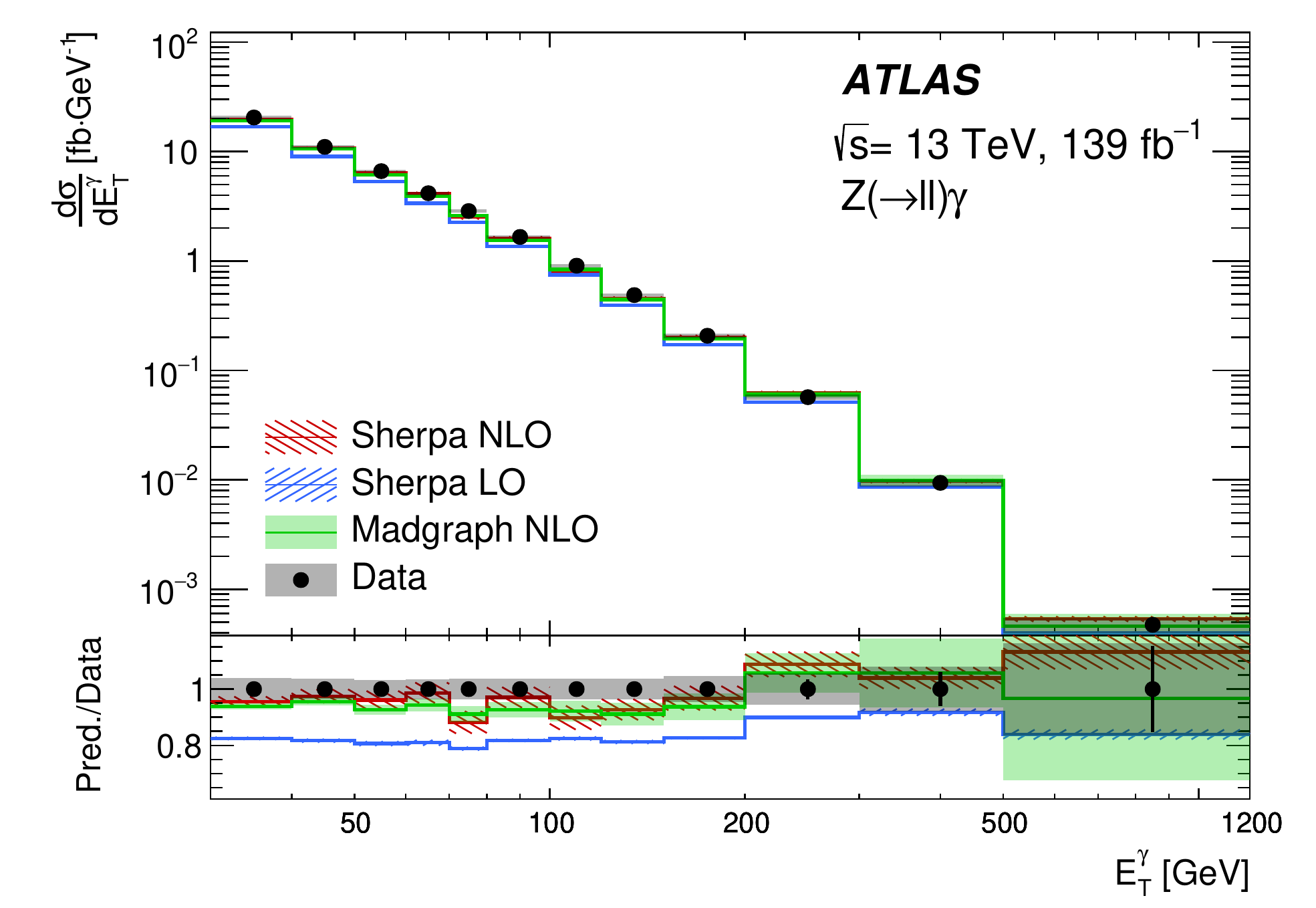}
\includegraphics[width=0.495\textwidth]{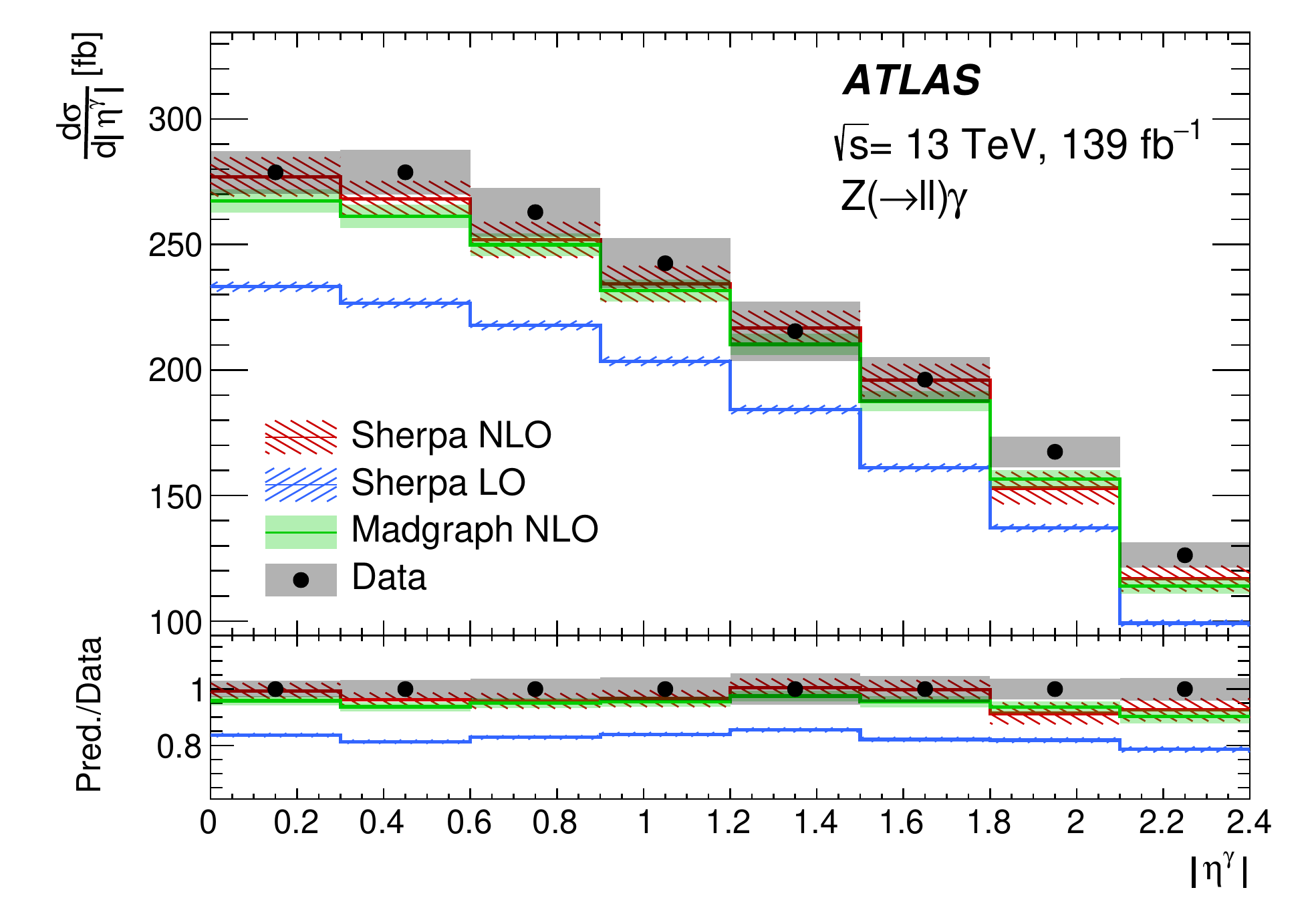}  \\
\includegraphics[width=0.495\textwidth]{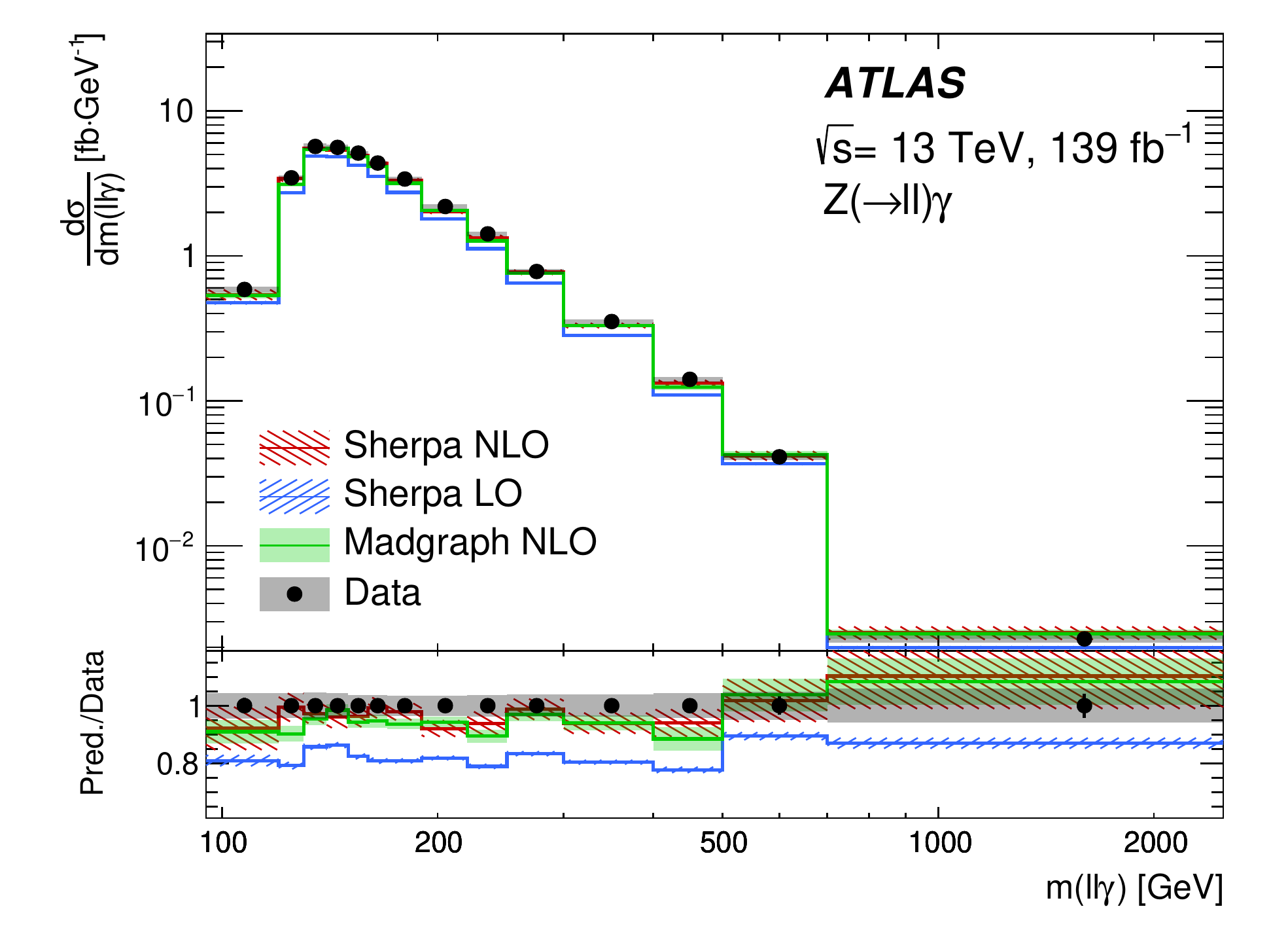}
\includegraphics[width=0.495\textwidth]{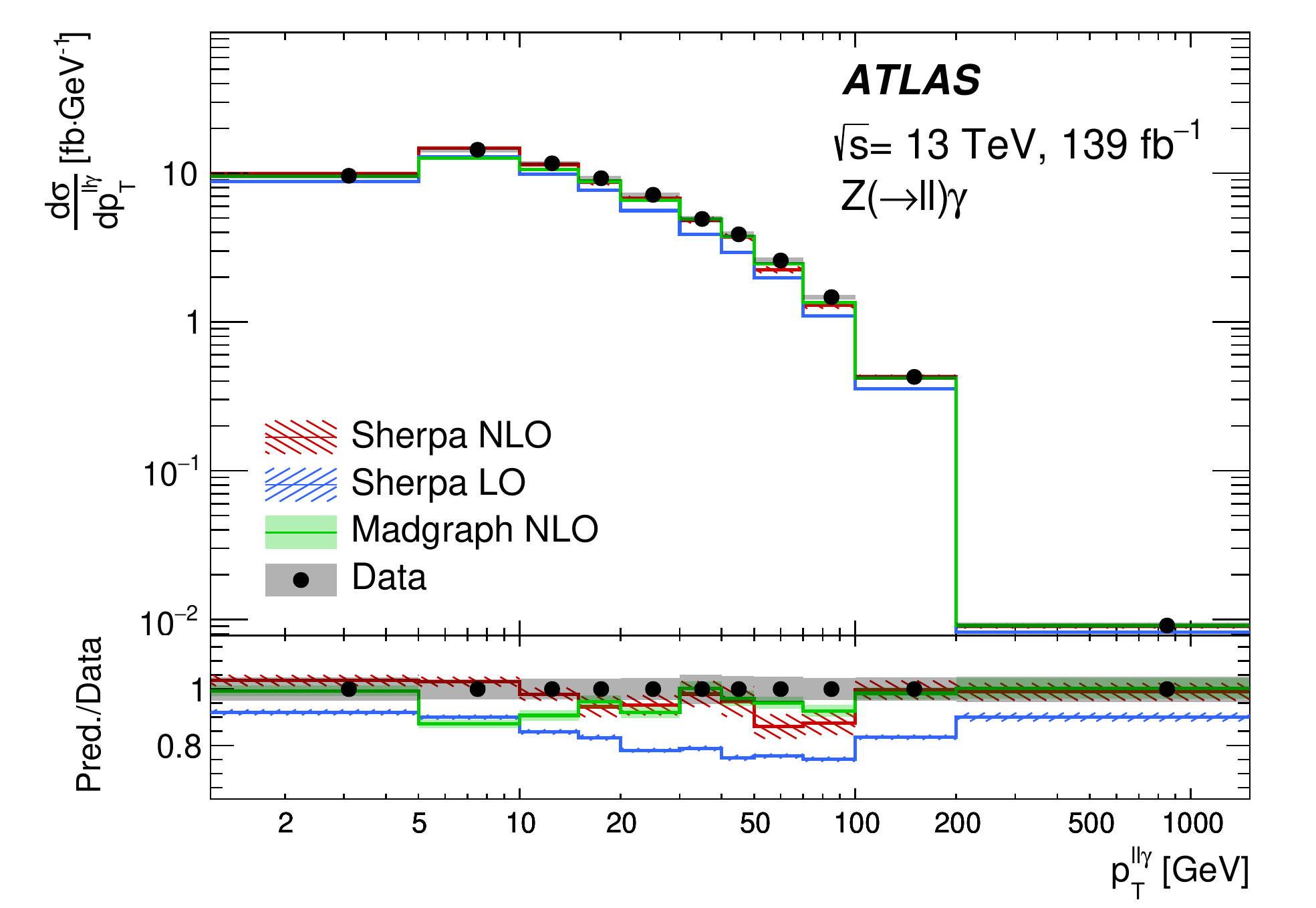} \\
\includegraphics[width=0.495\textwidth]{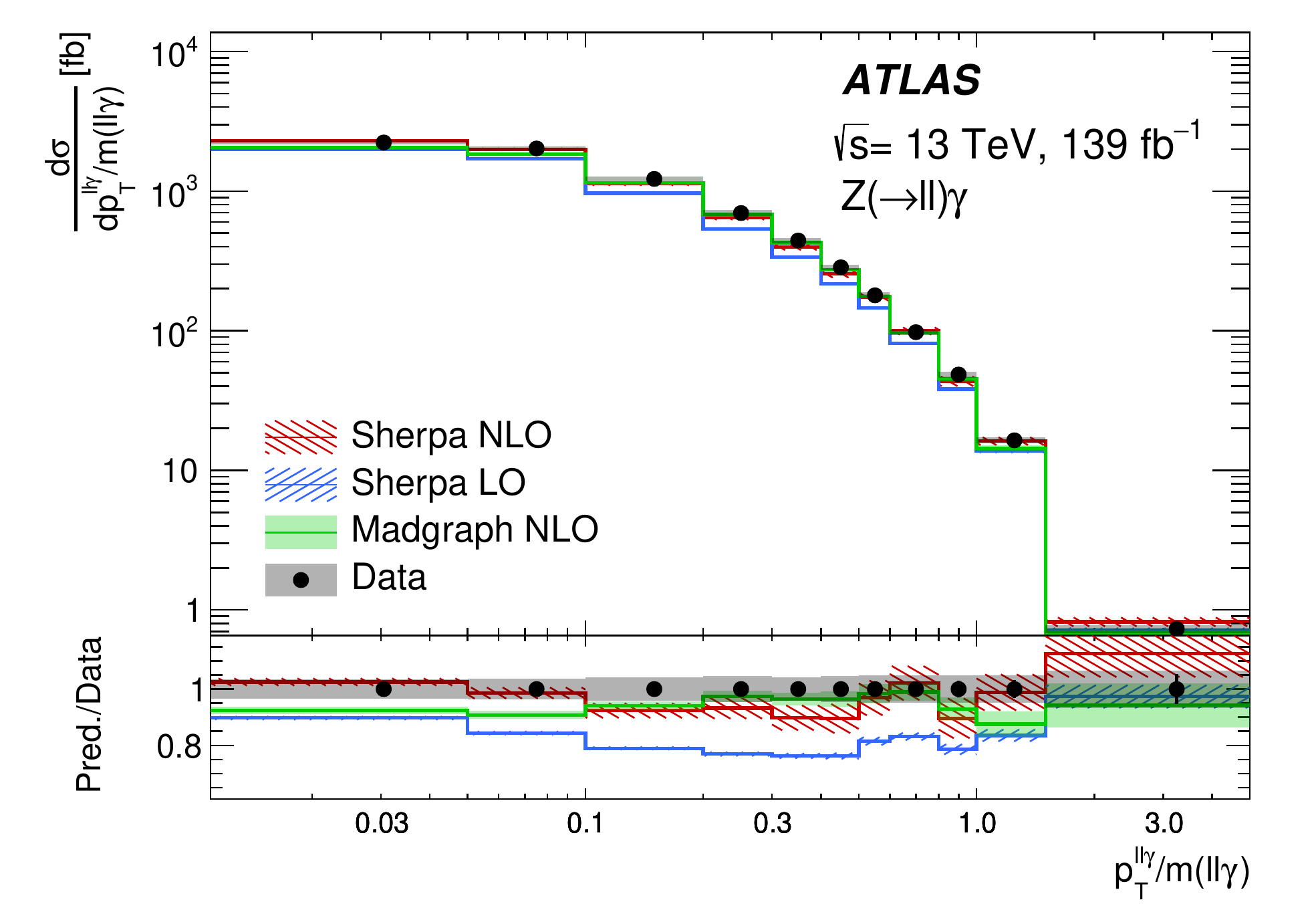}
\includegraphics[width=0.495\textwidth]{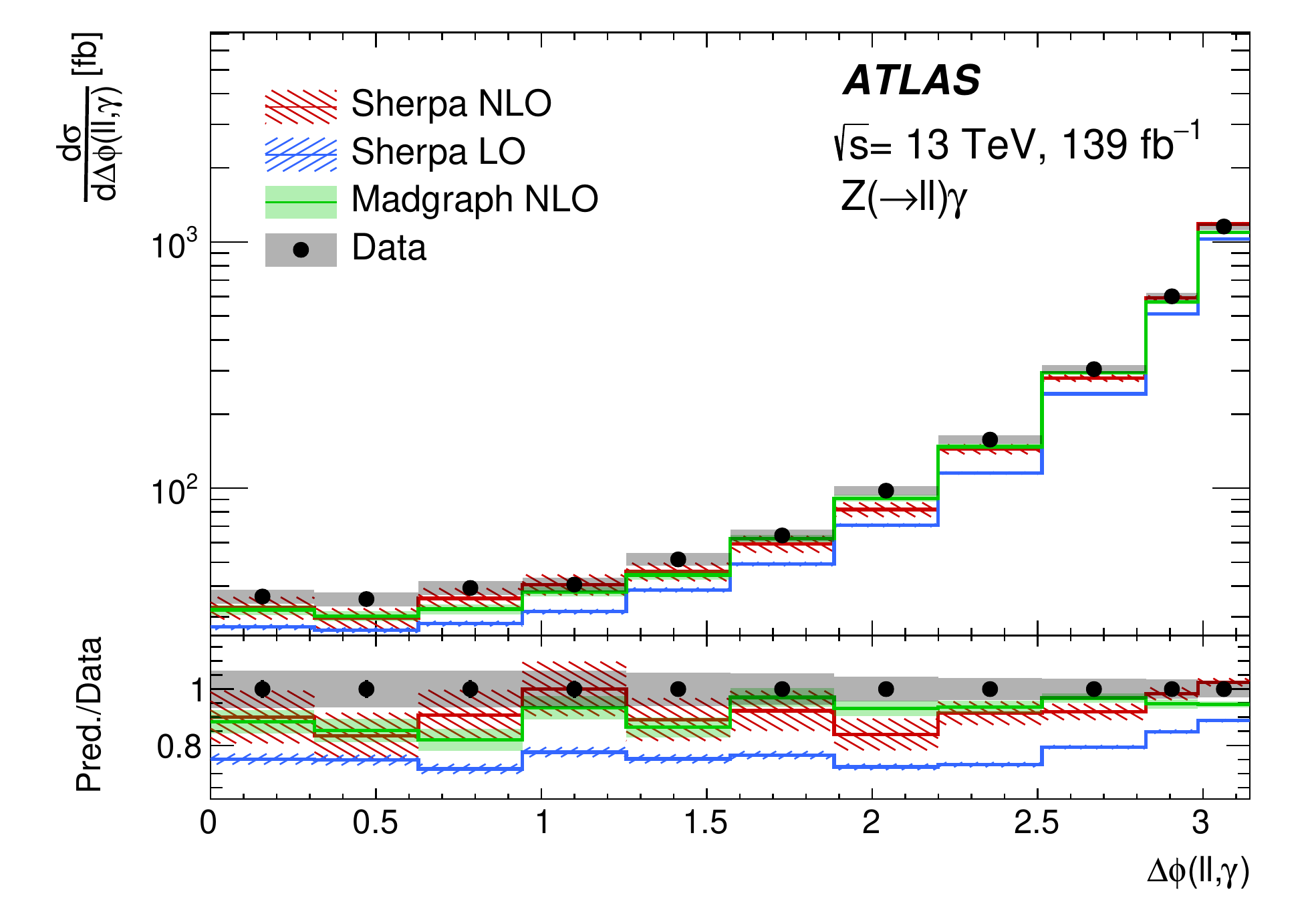}  \\
\caption{Measured fiducial cross-sections (black data points)
for the observables (from left to right and top to bottom) \etg, \aetag, \mllg, \ptllg, \rllg and \dphillg for the $pp \rightarrow Z(\ell^+\ell^-)\gamma$ process.
The error bars on the data points show the statistical uncertainty in the measured values.
The grey shaded regions show the total uncertainty in the unfolded data, excluding the uncertainty in the integrated luminosity.
The measured cross-sections are compared with SM expectations obtained from the \SHERPA and \MGMCatNLO event generators at particle level.
The uncertainty bands on the MC predictions show the statistical uncertainty due to the limited number of MC events.
The lower section of each plot shows the ratio of the SM expectation to the measured cross-section.}
\label{fig:UnfoldedDataVsSHERPA}
\end{center}
\end{figure}

\begin{figure}[hbtp]
\begin{center}
\includegraphics[width=0.495\textwidth]{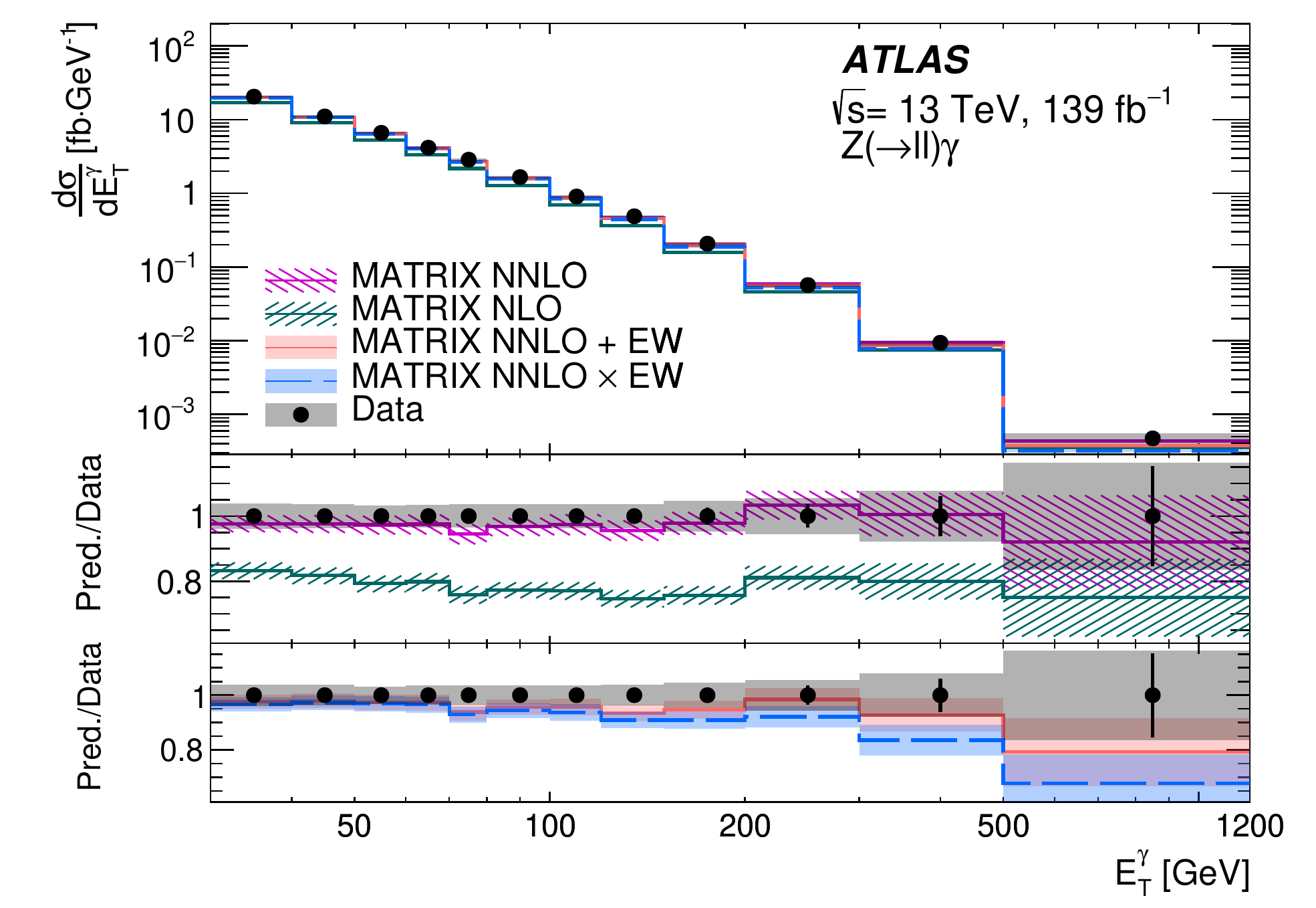}
\includegraphics[width=0.495\textwidth]{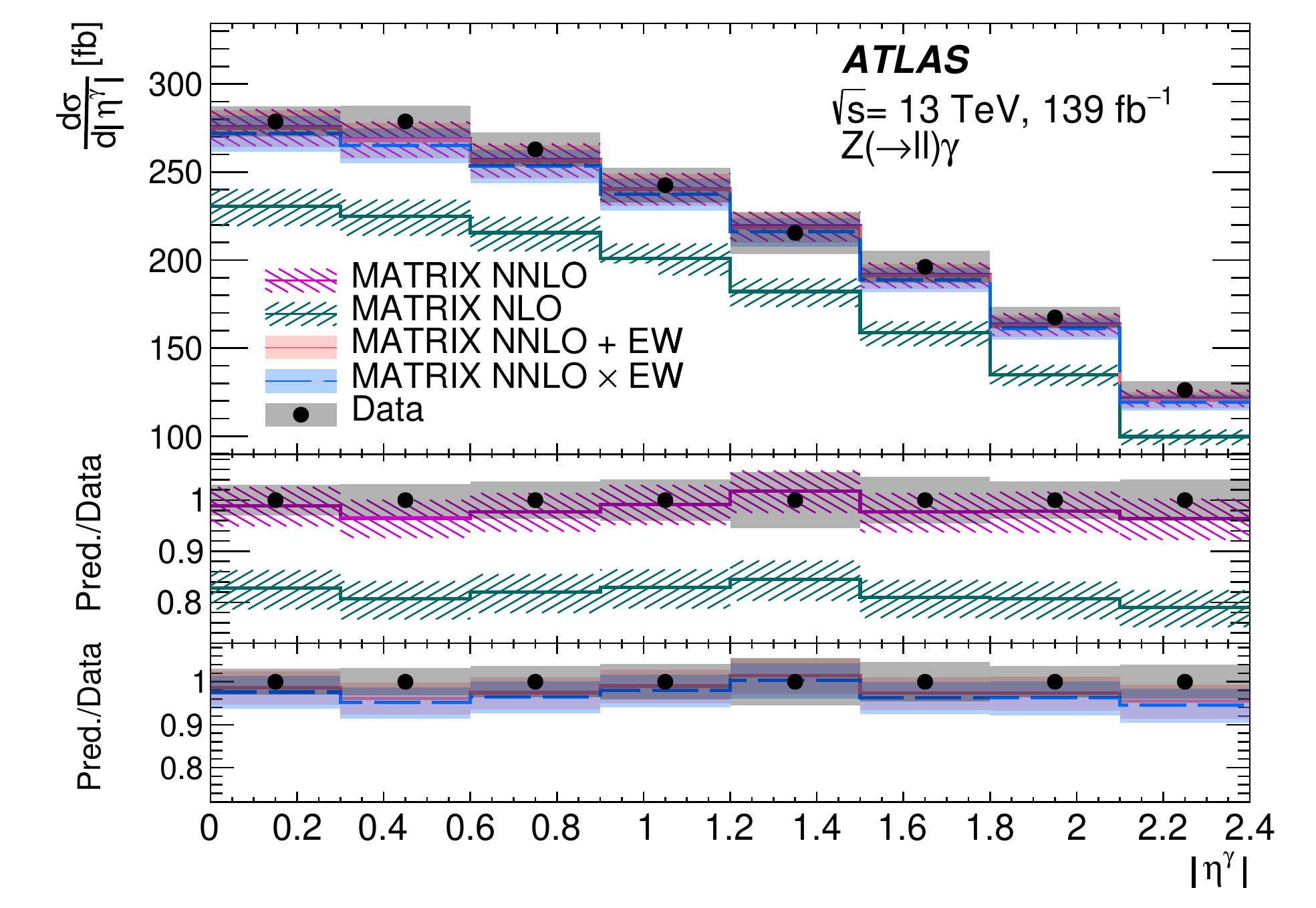}  \\
\includegraphics[width=0.495\textwidth]{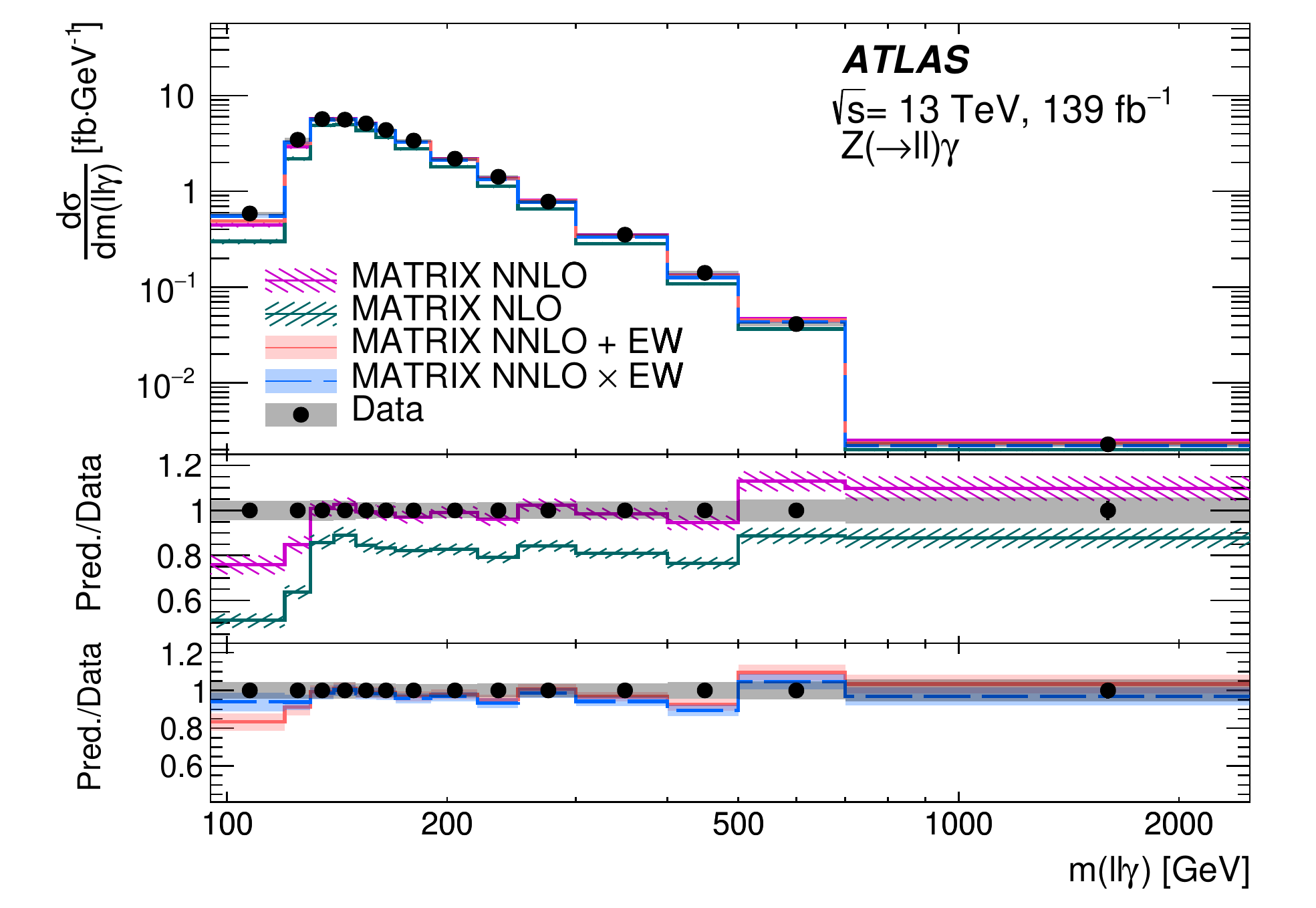}
\includegraphics[width=0.495\textwidth]{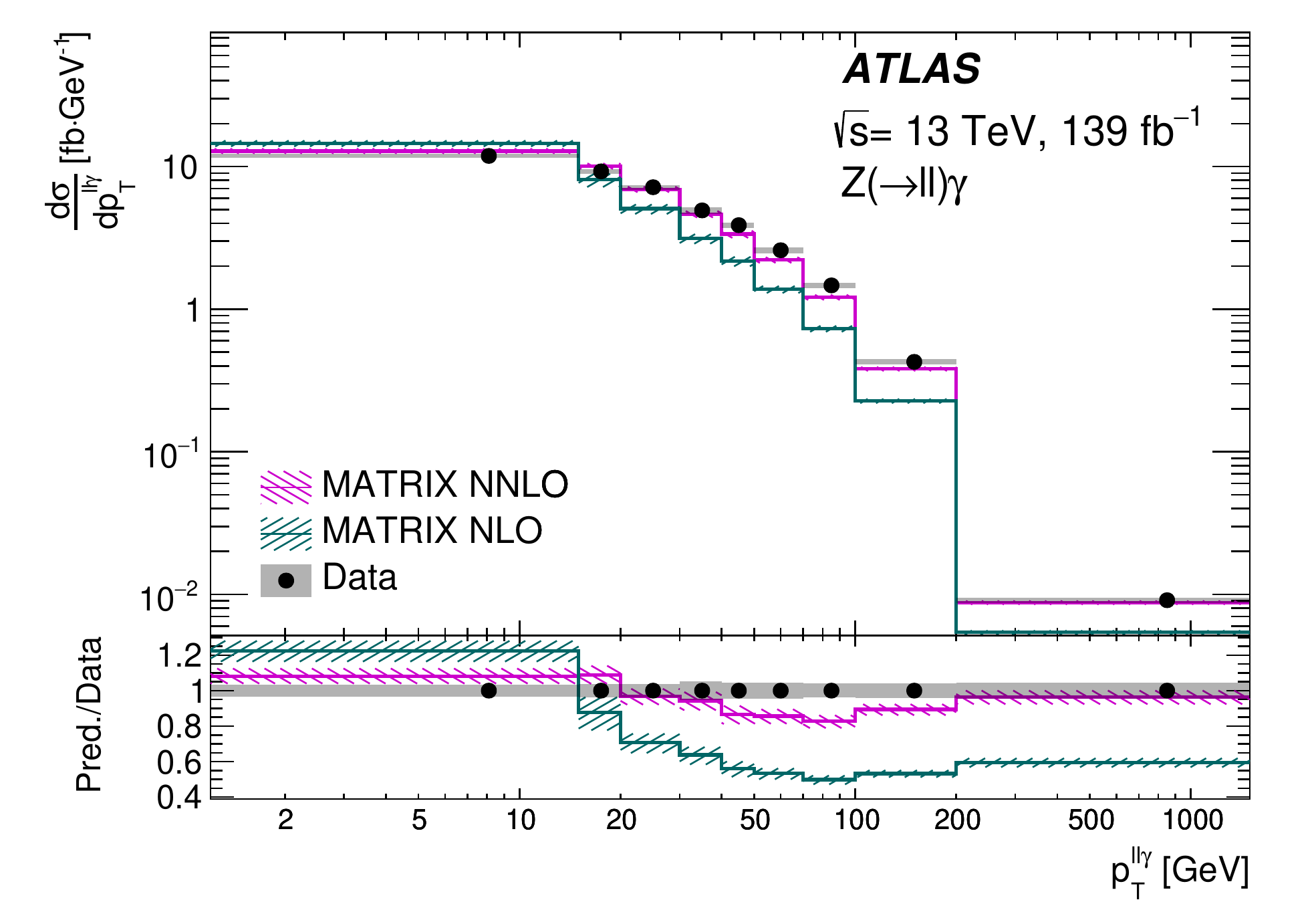}  \\
\includegraphics[width=0.495\textwidth]{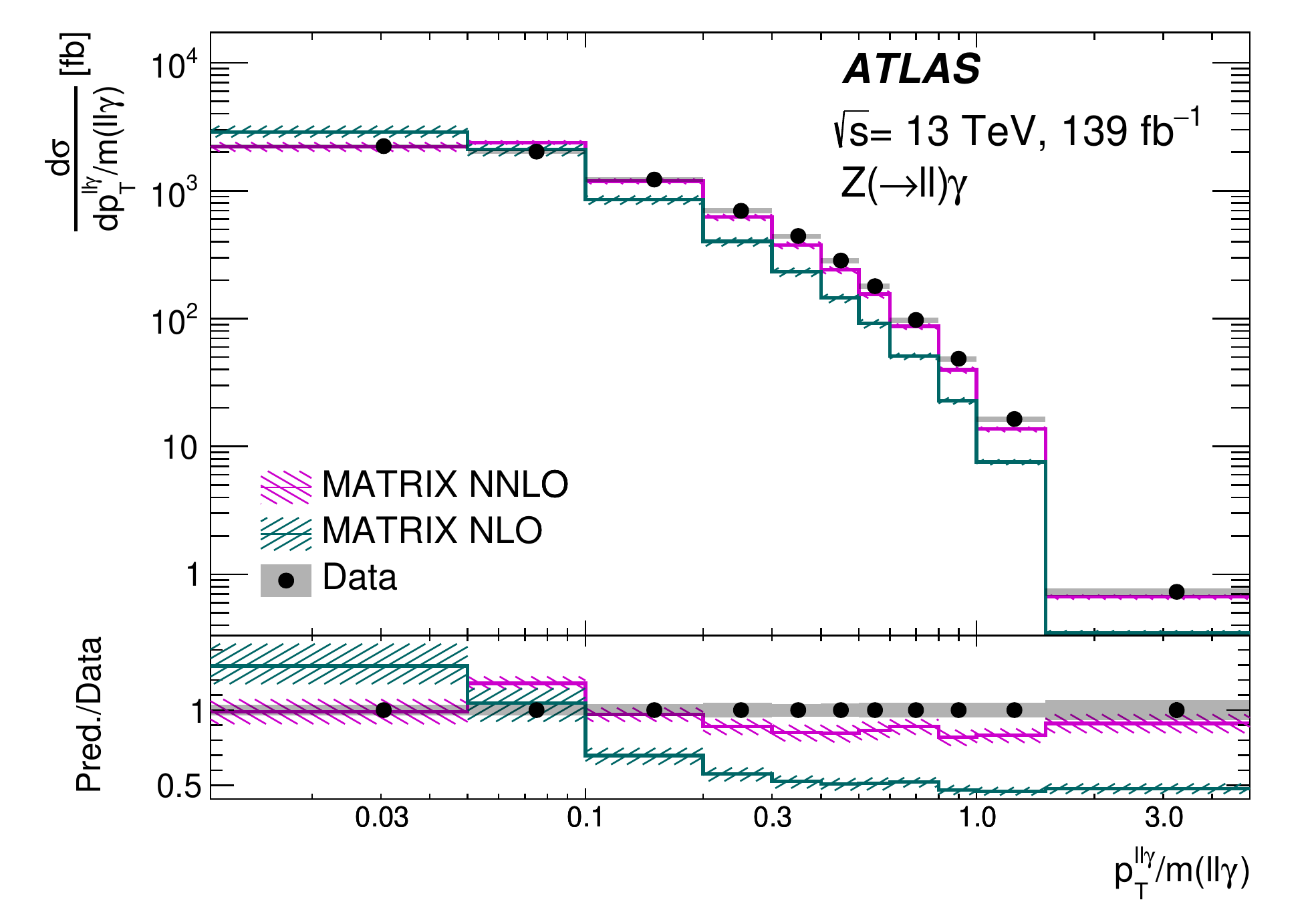}
\includegraphics[width=0.495\textwidth]{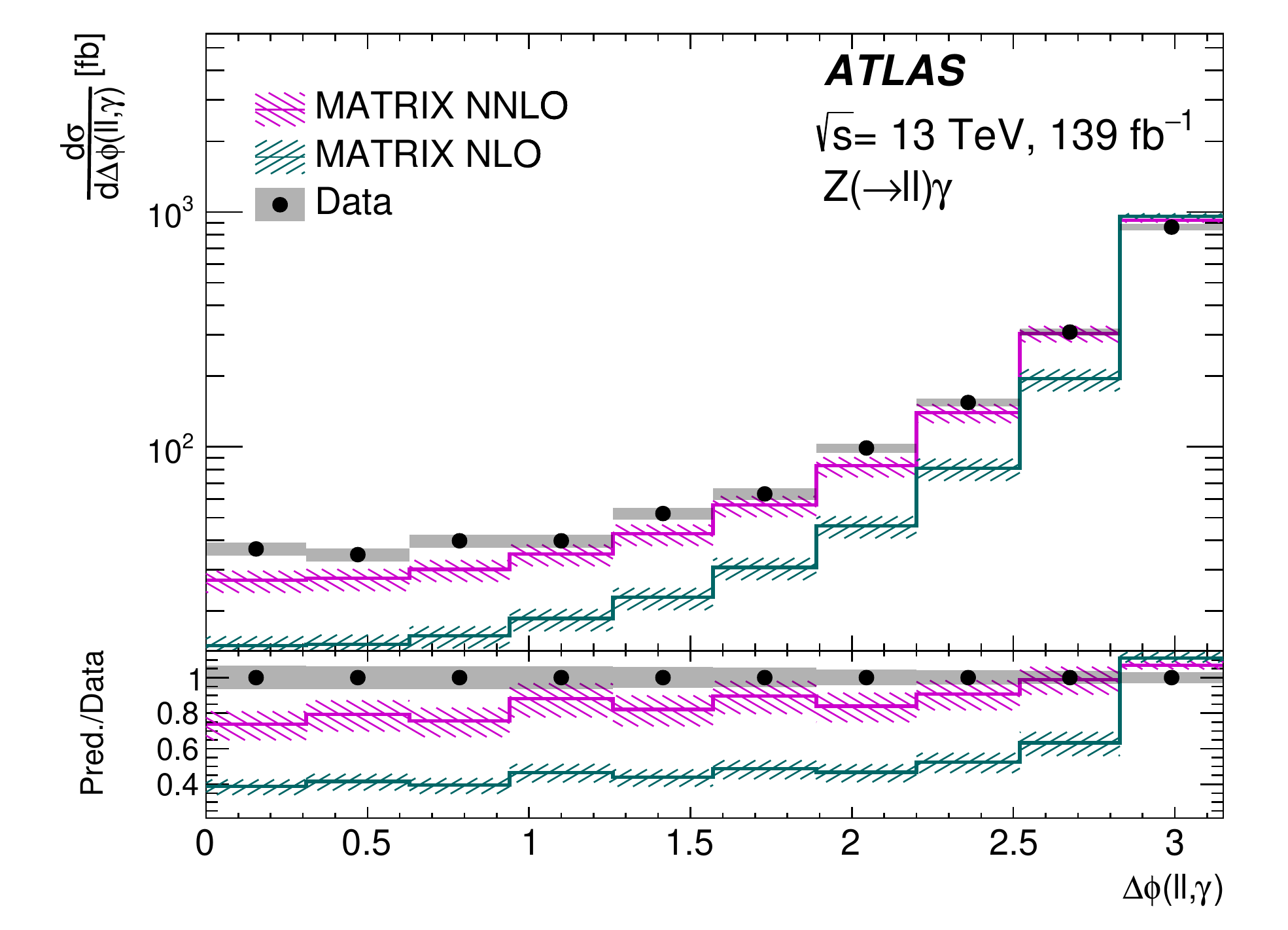}  \\
\caption{Measured fiducial cross-sections (black data points)
for the observables (from left to right and top to bottom) \etg, \aetag, \mllg, \ptllg, \rllg and \dphillg for the $pp \rightarrow Z(\ell^+\ell^-)\gamma$ process.
The measured cross-sections are compared with SM expectations obtained from the \MATRIX parton-level generator, corrected to particle level.
The error bars on the data points show the statistical uncertainty in the measured values.
The grey shaded regions show the total uncertainty in the unfolded data, excluding the uncertainty in the integrated luminosity.
The purple and green hatched regions show the total uncertainty in the \MATRIX predictions.
For the quantities \etg, \aetag and \mllg, the blue (orange) histograms show the \MATRIX NNLO cross-sections with EW NLO corrections applied multiplicatively (additively),
while the blue (orange) shaded regions show the corresponding total uncertainties.
The lower panel(s) in each plot show the ratio of the SM expectation to the measured cross-section.}
\label{fig:UnfoldedDataVsMATRIX}
\end{center}
\end{figure}
 
The SM expectations shown in \Fig{\ref{fig:UnfoldedDataVsSHERPA}} are obtained from parton shower MC samples, at LO and NLO, as described in \Sect{\ref{sec:SMpredictions}}.
The SM expectations shown in \Fig{\ref{fig:UnfoldedDataVsMATRIX}} are obtained from NLO and NNLO calculations at parton level,
with parton-to-particle corrections applied, again as described in \Sect{\ref{sec:SMpredictions}}.
For the \ptllg and \dphillg distributions, fixed-order calculations such as those carried out by \MATRIX
are not expected to describe the data well because of the importance of soft-gluon resummation effects.
To enable a comparison with the \MATRIX predictions, the first three bins in the \ptllg distribution of \Fig{\ref{fig:UnfoldedDataVsSHERPA}}, covering $\ptllg<15$~\GeV,
and the last two bins in the \dphillg distribution, covering $0.9\pi < \dphillg < \pi$,
are shown combined in \Fig{\ref{fig:UnfoldedDataVsMATRIX}}.
 
The predictions from \SHERPA at LO underestimate the measured rate by typically 10--25\%,
but give a generally good description of the shape of the observed kinematic distributions,
although clear differences are seen for \ptllg, \rllg and \dphillg.
The predicted rates and shapes from \SHERPA and \MGMCatNLO at NLO are in closer agreement with observation, although differences in shape persist for the \dphillg distribution.
The NLO prediction from \MATRIX generally underestimates the measured cross-section, especially at high \ptllg and \rllg, and at low \mllg and \dphillg,
where the disagreement with data can be as large as about 60\%.
Agreement between the \MATRIX prediction and data is much improved at NNLO, although the NNLO prediction continues to underestimate the measured cross-section in some regions of phase space,
especially in the region $\mllg<130$~\GeV, and for low values of \dphillg.
 
The effect of NLO EW corrections on the predicted differential cross-sections from \MATRIX at NNLO is shown in \Fig{\ref{fig:UnfoldedDataVsMATRIX}} for the observables \etg, \aetag and \mllg for which such corrections are available.
The corrected cross-sections are shown separately with the component of the EW corrections arising from partonic processes with a $q\overline{q}$ initial state
applied either multiplicatively or additively.
The EW corrections are negative in all bins of the measured differential cross-sections, except for the lowest two bins in \mllg.
They are largest (and negative) at high \etg, where they become of similar order to the difference between the predicted cross-sections from \MATRIX computed at NLO and NNLO in pQCD.
 
The SM expectations shown in \Figs{\ref{fig:UnfoldedDataVsSHERPA}}{\ref{fig:UnfoldedDataVsMATRIX}} include the contribution from EW \zgjj production,
obtained as described in \Sect{\ref{sec:SMpredictions}}.
The largest relative contribution from this process
is predicted to arise for the highest bins of \etg and \ptllg, where it reaches about 8\% of the \SHERPA LO prediction.
 
\FloatBarrier
\section{Summary}
\label{sec:summary}
The cross-section for the production of a \Zboson boson in association with a high-energy prompt photon is measured using 139\,\ifb\ of proton--proton collision data at $\sqrt{s} = 13$~\TeV\
collected with the ATLAS detector at the LHC.
The analysis selects events in the \epemg and \mpmmg channels, and is performed in a phase-space region
defined by kinematic requirements on the leptons and the photon, and by requiring the photon to be isolated.
 
Differential cross-sections are presented as functions of the transverse energy and pseudorapidity of the photon,
and as functions of the transverse momentum and invariant mass of the \lplmg system, their ratio,
and the angle between the transverse directions of the lepton pair and the photon.
 
The results are compared with SM expectations derived from the parton shower Monte Carlo event generators \SHERPA, at LO and NLO in pQCD, \MGMCatNLO at NLO,
and from the parton-level generator \MATRIX, corrected to particle level, at NLO and NNLO.
The integrated fiducial-region cross-sections predicted by \MGMCatNLO at NLO, \SHERPA at NLO, and by \MATRIX at NNLO
underestimate the measured cross-section by about 6\%, 4\% and 3\%, respectively,
but are in agreement with measurement within the uncertainties.
The corresponding predictions for the shapes of the kinematic distributions describing the \lplmg system are  generally in good agreement with observation,
although some differences are seen, especially for the \MATRIX NNLO prediction at low \mllg and low \dphillg.
 
The precision achieved for the measurement of the inclusive cross-section is 2.9\%,
representing about a factor of two improvement compared with the corresponding ATLAS result at $\sqrt{s}=8$~TeV.

\section*{Acknowledgements}

We thank CERN for the very successful operation of the LHC, as well as the
support staff from our institutions without whom ATLAS could not be
operated efficiently.
 
We acknowledge the support of ANPCyT, Argentina; YerPhI, Armenia; ARC, Australia; BMWFW and FWF, Austria; ANAS, Azerbaijan; SSTC, Belarus; CNPq and FAPESP, Brazil; NSERC, NRC and CFI, Canada; CERN; CONICYT, Chile; CAS, MOST and NSFC, China; COLCIENCIAS, Colombia; MSMT CR, MPO CR and VSC CR, Czech Republic; DNRF and DNSRC, Denmark; IN2P3-CNRS and CEA-DRF/IRFU, France; SRNSFG, Georgia; BMBF, HGF and MPG, Germany; GSRT, Greece; RGC and Hong Kong SAR, China; ISF and Benoziyo Center, Israel; INFN, Italy; MEXT and JSPS, Japan; CNRST, Morocco; NWO, Netherlands; RCN, Norway; MNiSW and NCN, Poland; FCT, Portugal; MNE/IFA, Romania; MES of Russia and NRC KI, Russia Federation; JINR; MESTD, Serbia; MSSR, Slovakia; ARRS and MIZ\v{S}, Slovenia; DST/NRF, South Africa; MINECO, Spain; SRC and Wallenberg Foundation, Sweden; SERI, SNSF and Cantons of Bern and Geneva, Switzerland; MOST, Taiwan; TAEK, Turkey; STFC, United Kingdom; DOE and NSF, United States of America. In addition, individual groups and members have received support from BCKDF, CANARIE, Compute Canada and CRC, Canada; ERC, ERDF, Horizon 2020, Marie Sk{\l}odowska-Curie Actions and COST, European Union; Investissements d'Avenir Labex, Investissements d'Avenir Idex and ANR, France; DFG and AvH Foundation, Germany; Herakleitos, Thales and Aristeia programmes co-financed by EU-ESF and the Greek NSRF, Greece; BSF-NSF and GIF, Israel; CERCA Programme Generalitat de Catalunya and PROMETEO Programme Generalitat Valenciana, Spain; G\"{o}ran Gustafssons Stiftelse, Sweden; The Royal Society and Leverhulme Trust, United Kingdom.
 
The crucial computing support from all WLCG partners is acknowledged gratefully, in particular from CERN, the ATLAS Tier-1 facilities at TRIUMF (Canada), NDGF (Denmark, Norway, Sweden), CC-IN2P3 (France), KIT/GridKA (Germany), INFN-CNAF (Italy), NL-T1 (Netherlands), PIC (Spain), ASGC (Taiwan), RAL (UK) and BNL (USA), the Tier-2 facilities worldwide and large non-WLCG resource providers. Major contributors of computing resources are listed in Ref.~\cite{ATL-GEN-PUB-2016-002}.
 
 
\clearpage
\printbibliography
 
\clearpage
 
\begin{flushleft}
{\Large The ATLAS Collaboration}

\bigskip

G.~Aad$^\textrm{\scriptsize 102}$,    
B.~Abbott$^\textrm{\scriptsize 129}$,    
D.C.~Abbott$^\textrm{\scriptsize 103}$,    
A.~Abed~Abud$^\textrm{\scriptsize 71a,71b}$,    
K.~Abeling$^\textrm{\scriptsize 53}$,    
D.K.~Abhayasinghe$^\textrm{\scriptsize 94}$,    
S.H.~Abidi$^\textrm{\scriptsize 167}$,    
O.S.~AbouZeid$^\textrm{\scriptsize 40}$,    
N.L.~Abraham$^\textrm{\scriptsize 156}$,    
H.~Abramowicz$^\textrm{\scriptsize 161}$
% ...
\bigskip
\\

% ...
$^{40}$Niels Bohr Institute, University of Copenhagen, Copenhagen; Denmark.\\
% ...
$^{53}$II. Physikalisches Institut, Georg-August-Universit\"{a}t G\"ottingen, G\"ottingen; Germany.\\
% ...
$^{71}$$^{(a)}$INFN Sezione di Pavia;$^{(b)}$Dipartimento di Fisica, Universit\`a di Pavia, Pavia; Italy.\\
% ...
$^{94}$Department of Physics, Royal Holloway University of London, Egham; United Kingdom.\\
% ...
$^{102}$CPPM, Aix-Marseille Universit\'e, CNRS/IN2P3, Marseille; France.\\
$^{103}$Department of Physics, University of Massachusetts, Amherst MA; United States of America.\\
% ...
$^{129}$Homer L. Dodge Department of Physics and Astronomy, University of Oklahoma, Norman OK; United States of America.\\
% ...
$^{156}$Department of Physics and Astronomy, University of Sussex, Brighton; United Kingdom.\\
% ...
$^{161}$Raymond and Beverly Sackler School of Physics and Astronomy, Tel Aviv University, Tel Aviv; Israel.\\
% ...
$^{167}$Department of Physics, University of Toronto, Toronto ON; Canada.\\
% ...
$^{a}$ Also at Borough of Manhattan Community College, City University of New York, New York NY; United States of America.\\
$^{b}$ Also at CERN, Geneva; Switzerland.\\
% ...

\end{flushleft}


\clearpage
 
\end{document}